\documentclass[accepted]{melba}

\usepackage{amsmath,amsfonts}

\usepackage{amsmath,amsfonts}
\usepackage{marginnote}

\usepackage{amssymb}
\usepackage{booktabs}
\usepackage[table]{xcolor}
\usepackage{subcaption}

\usepackage{graphicx}
\usepackage{framed,multirow}
\usepackage{float}
\usepackage{amssymb}
\usepackage{rotating}
\usepackage{amsmath}
\usepackage{latexsym}
\usepackage{xurl}
\usepackage{url}
\usepackage{pifont}    
\usepackage{textcomp}  
\usepackage{xcolor}
\usepackage{xspace}
\usepackage{pdflscape}
\usepackage{placeins}
\usepackage{hyperref}
\usepackage{hyphenat}
\usepackage[normalem]{ulem}
\usepackage{tabularray}

\definecolor{myblue}{HTML}{3171AD}   %
\definecolor{mygreen}{HTML}{469C76}   %
\definecolor{myorange}{HTML}{D39334}   

\newcommand{\synthseg}{\texttt{Synth\discretionary{-}{}{}Seg}\xspace}
\newcommand{\frs}{\texttt{FetalReal\discretionary{-}{}{}Seg}\xspace}
\newcommand{\hrs}{\texttt{Real\discretionary{-}{}{}SynthHybrid}\xspace}
\newcommand{\fetseg}{\texttt{Fetal\discretionary{-}{}{}Synth\discretionary{-}{}{}Seg}\xspace}

\newcommand{\fabian}{\texttt{Fa\discretionary{-}{}{}Bi\discretionary{-}{}{}AN}\xspace}
\newcommand{\randfabian}{\texttt{rand\discretionary{-}{}{}Fa\discretionary{-}{}{}Bi\discretionary{-}{}{}AN}\xspace}
\newtheorem{hypothesis}{Hypothesis}

\melbaid{2026:023}  
\doi{10.59275/j.melba.2026-6e4e}
\melbaauthors{Vladyslav Zalevskyi and Thomas Sanchez}  
\email{vladyslav.zalevskyi@unil.ch}
\volume{3}
\firstpageno{482}  
\melbayear{2026}  
\datesubmitted{2025-12-02}  
\datepublished{2026-07-01}  

\ShortHeadings{Evaluating Synthetic Data Generation for Domain Generalization in Fetal Brain MRI Segmentation}{Vladyslav Zalevskyi and Thomas Sanchez}

\title{Evaluating Synthetic Data Generation for Domain Generalization in Fetal Brain MRI Segmentation}

\author{%
  \name \firstname Vladyslav \surname Zalevskyi\aff{1,2,*}\orcid{0009-0004-0191-299X},
  \name \firstname Thomas \surname Sanchez\aff{1,2,*}\orcid{0000-0003-3668-5155},
  \name \firstname Margaux \surname Roulet\aff{1,2}\orcid{0009-0005-9983-5706},
  \name \firstname Busra \surname Bulut\aff{1,2}\orcid{0009-0003-7787-7820},
  \name \firstname Hélène \surname Lajous\aff{1,2}\orcid{0000-0001-7729-6274},
  \name \firstname Jordina \surname Aviles Verdera\aff{3,4}\orcid{0009-0007-7575-4244},
  \name \firstname Sara \surname Neves Silva\aff{4}\orcid{0009-0009-7520-081X},
  \name \firstname Georg \surname Langs\aff{6,7,8}\orcid{0000-0002-5536-6873},
  \name \firstname Gregor \surname Kasprian\aff{7,9}\orcid{0000-0003-3858-3347},
  \name \firstname Roxane \surname Licandro\aff{6,7,8}\orcid{0000-0001-9066-4473},
  \name \firstname Jana \surname Hutter\aff{3,4}\orcid{0000-0003-3476-3500},
  \name \firstname Hamza \surname Kebiri\aff{1,2}\orcid{0000-0001-7592-3166},
  \name \firstname Meritxell \surname Bach Cuadra\aff{1,2}\orcid{0000-0003-2730-4285}
}

\affiliations{%
  \num 1 \addr Department of Radiology, Lausanne University Hospital and University of Lausanne (UNIL), Lausanne, Switzerland \\
  \num 2 \addr CIBM Center for Biomedical Imaging, Lausanne, Switzerland \\
  \num 3 \addr Institute for Information Processing, Leibniz University Hannover, Hannover, Germany \\
  \num 4 \addr Department of Biomedical Engineering, School of Biomedical Engineering \& Imaging Sciences, King’s College London, London, United Kingdom \\
  \num 5 \addr Department of Biomedical Imaging and Image-Guided Therapy, Division of Neuroradiology and Musculoskeletal Radiology, Medical University of Vienna, Vienna, Austria \\
  \num 6 \addr Department of Biomedical Imaging and Image-guided Therapy, Computational Imaging Research Lab (CIR), Medical University of Vienna, Vienna, Austria\\
   \num 7 \addr Christian Doppler Laboratory for Mathematical Modelling and Simulation of Next-Generation Medical Ultrasound Devices, Medical University of Vienna, Vienna, Austria\\
    \num 8 \addr Comprehensive Center for Artificial Intelligence in Medicine, Medical University of Vienna, Vienna, Austria\\
 \num 9 \addr Division of Neuroradiology and Musculoskeletal Radiology, Department of Biomedical Imaging and Image–guided Therapy, Medical University of Vienna, Vienna, Austria \\
  \num * \addr Equal contribution
}

\abstract{
	Fetal brain tissue segmentation from magnetic resonance imaging (MRI) is crucial for studying neurodevelopment, but remains challenging due to data heterogeneity and limited annotations. Domain randomization (DR) has recently emerged as a promising strategy for single-source domain generalization by synthesizing training images with randomized artifacts, contrast, and resolution.
\textcolor{black}{In this work, we investigate how to maximize the out-of-domain (OOD) generalization of DR-based methods. We evaluate several synthetic data generation strategies for DR, with a particular focus on our recently proposed framework, \fetseg. We show that simple Gaussian mixture-based intensity modeling outperforms more complex physics-based simulations, and that \textit{intensity clustering} (subdividing tissue classes based on intensity) improves OOD robustness.}
Evaluated on 348 fetal subjects from four sites spanning 0.55–3T and both T1w and T2w contrasts, \fetseg reaches state-of-the-art performance on several FeTA 2024 testing datasets (80–85 Dice score) and, for the first time, offers robust segmentation on modalities other than T2w for fetal brain segmentation (80 Dice on dHCP-T1w dataset). Compared with state-of-the-art methods such as BOUNTI, nnU-Net ensemble, and the FeTA 2024 winner, \fetseg delivers comparable or superior accuracy while maintaining strong robustness across domain shifts.
	Our code, model weights, and Docker image ready for easy inference are available at~\url{https://hub.docker.com/r/vzalevskyi/fetalsynthseg}.}


\begin{document}

\twocolumn[\maketitle]

\begin{figure*}[t]
    \centering
    \includegraphics[width=1\linewidth]{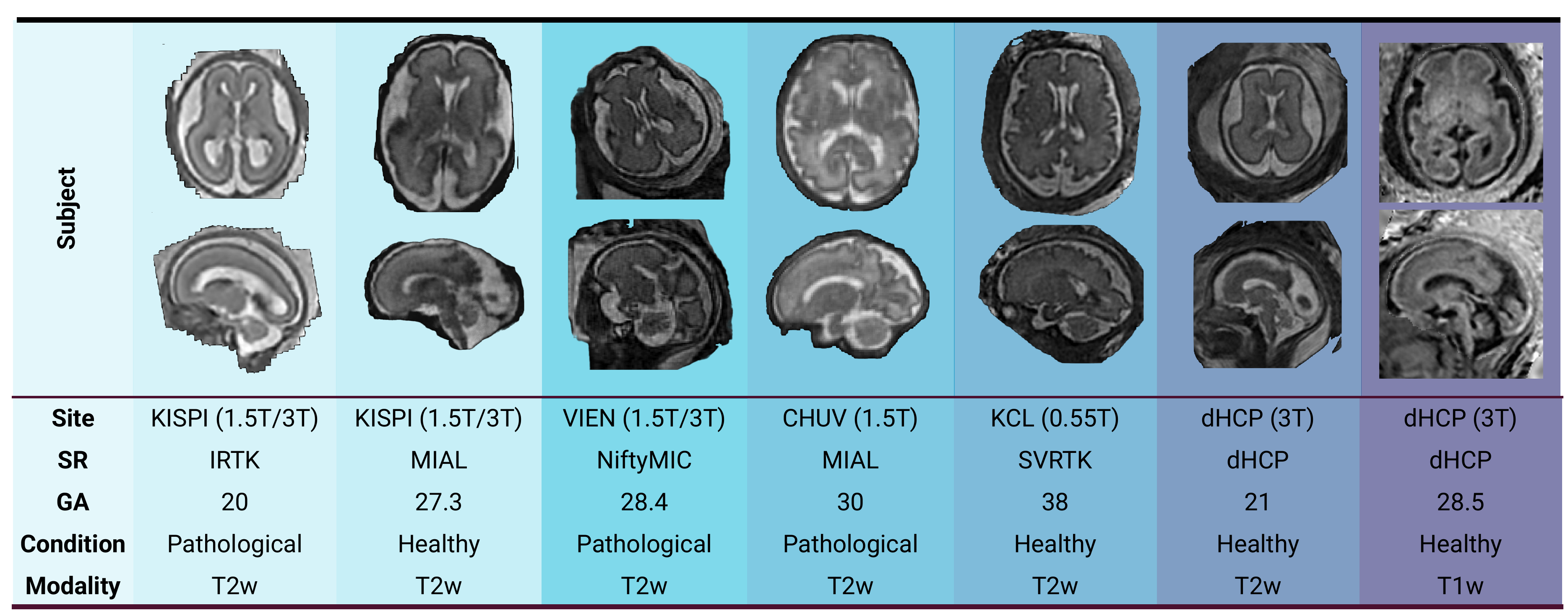}
    \vspace{-.2cm}
    \caption{\textbf{Domain shifts are ubiquitous in fetal brain super-resolution MRI.} Variations in acquisition protocols, super-resolution methods, gestational age and pathology distributions, field strengths, and specific MRI modalities contribute to distributional differences that affect model generalization.}
    
    \label{fig:domain_shifts_ex}
\end{figure*}

\section{Introduction}
Fetal development is a critical period shaping lifelong neurological and physiological outcomes~\citep{Halfon2014-bp}. Monitoring brain maturation and detecting atypical development~during this period enables early diagnosis and intervention \citep{SALEEM2014507}. While ultrasonography (US) remains the default fetal imaging modality because of its broad availability and its low cost, fetal magnetic resonance imaging (MRI) complements US: it provides superior soft-tissue contrast and is less operator-dependent~\citep{Glenn1604}. MRI has been used to study fetal brain development~\citep{Jakab2021, AvilesVerdera2023, Machado2022, Shen2022} and for the detection of abnormalities such as ventriculomegaly, spina bifida, corpus callosum agenesis, edema, and hemorrhage~\citep{garel2004mri,Pfeifer2019}. Although, automated methods for fetal brain MRI analysis are promising to enable rapid and reliable scan processing~\citep{Uus2023.04.18.537347_Bounti}, they struggle with limited data availability and domain shifts, which are particularly pronounced in the fetal population~\citep{dockes2021preventing,varoquaux2022machine,Guan2022}. 

\looseness=-1
Domain shifts, or distribution shifts --- a mismatch between the distribution of training and testing data --- are quite severe in MRI and stem from differences in scanner hardware, imaging protocols, and reconstruction methods \citep{Yan2020}. Fetal MRI exacerbates these challenges as the fetal brain morphology heavily changes during gestation and can be drastically altered by pathological processes~\citep{Stiles2010, Dubois2020}. Furthermore, fetal MR images are typically acquired as T2-weighted orthogonal stacks of two-dimensional (2D) slices using fast spin echo sequences to mitigate motion artifacts. Post-processing using super-resolution (SR) reconstruction algorithms is then needed to create a high-resolution three-dimensional (3D) volume~\citep{Rousseau2010,gholipour2010robust,kuklisova-murgasova_reconstruction_2012,tourbier_efficient_2015, ebner_automated_2020},  which induces an additional layer of heterogeneity. As a result, training models that generalize across these dimensions (Figure~\ref{fig:domain_shifts_ex}) remains challenging~\citep{payette2024multicenter,zalevskyi2025advancesautomatedfetalbrain}.

Many strategies have been proposed to mitigate domain shifts, including diffusion-based models \citep{kaandorp2025pathologicalmrisegmentationsynthetic,niemeijer2024generalization} and data augmentation-based approaches \citep{shorten2019survey}. Recently, domain randomization~\citep{tobin2017domain} techniques have seen great success in the field of MR image analysis~\citep{billot2021synthseg,billot2023robust,Liu_2023_BrainID}. These methods begin with label maps rather than real images, generating synthetic images with randomized contrasts using Gaussian mixture models, alongside common artifact corruptions, and demonstrated excellent OOD generalization performance, especially in MRI~\citep{billot2023robust}.

\paragraph{\textbf{Contributions}}
This work extends our MICCAI 2024 publication~\citep{zalevskyi2024improving} by expanding the evaluation of the proposed \fetseg model to additional datasets and presenting new ablation studies and comparisons with state-of-the-art (SoTA) methods. Specifically, in this work, we aim to investigate two main research hypotheses:
\vspace{-.1cm}
\begin{hypothesis}\label{hyp:h1}
Do domain randomization methods, by simulating a wide range of contrasts, provide more effective data augmentation than physics-based approaches?
\end{hypothesis}

\begin{figure*}[t]
    \centering
    \includegraphics[width=1.0\linewidth]{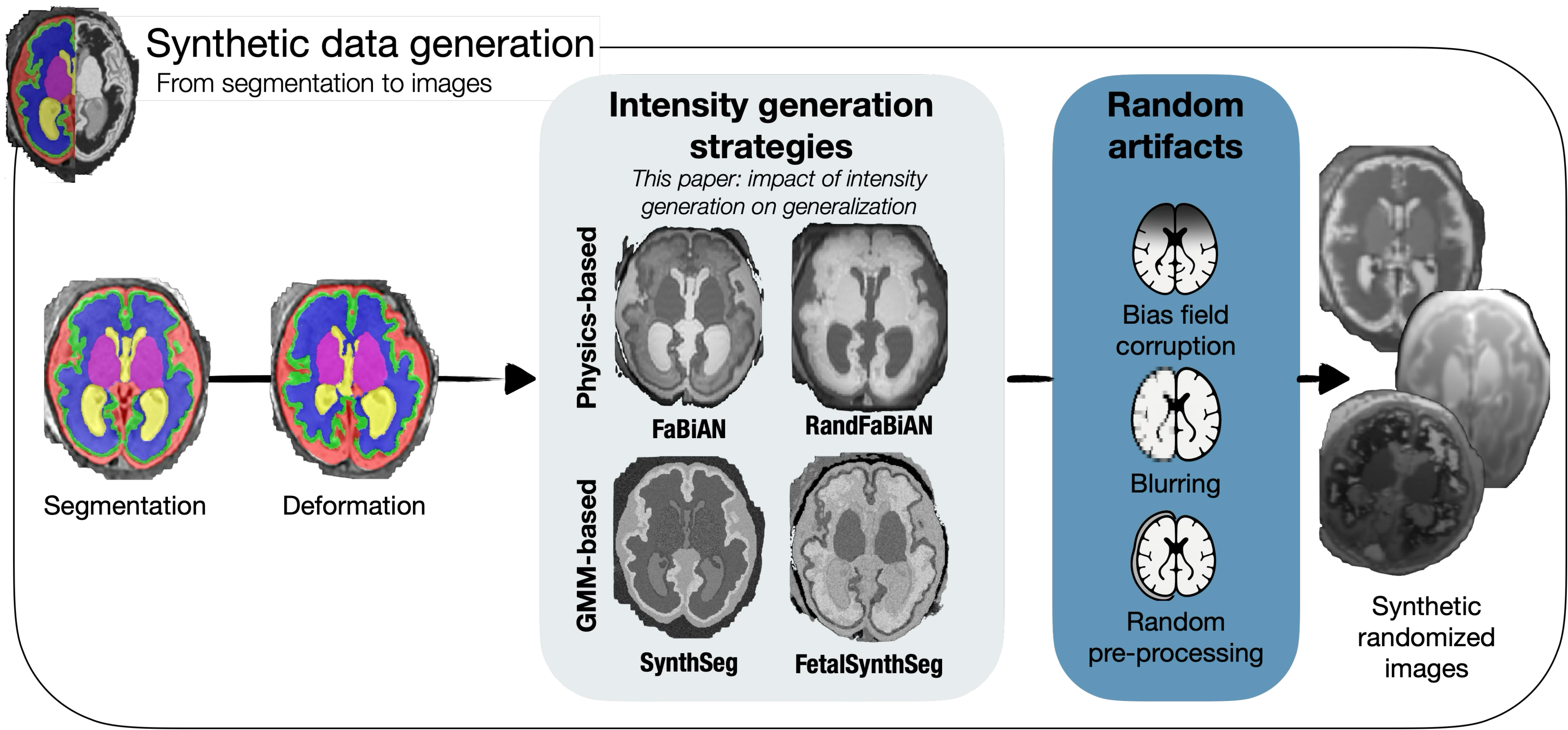}
    \caption{\textbf{Summary of the contribution of this paper:} we systematically investigate how strategies for intensity generation impact the OOD generalization and show that our proposed method, \fetseg, achieves strong generalization ability across sites, reconstruction methods and imaging modalities.}
    \label{fig:graphical}
\end{figure*}

\begin{hypothesis}\label{hyp:h2}
Do domain randomization methods enable more robust OOD generalization than models trained using only real data, while maintaining competitive in-domain (ID) performance?
\end{hypothesis}
Our results explore and confirm both hypotheses, and also disentangle which components of the data generation pipeline contribute most to out-of-domain robustness, what types of synthetic contrast modeling are most effective, and under which conditions such approaches may fail when applied to diverse clinical and research cohorts. In this work, we focus on the data synthesis and augmentation strategies rather than the design of new neural network architectures, while the main methodological contributions lie in the non-learning-based components that generate and randomize the training data. By isolating and  evaluating these components, our study aims to provide insights that remain applicable across future generations of segmentation networks. We summarize our key contributions are as follows:

\noindent\textbf{(i)}  We compare \textbf{Gaussian mixture-based} contrast simulation with \textbf{physics-based} approaches to assess the benefits and drawbacks of physically grounded modeling (Figure \ref{fig:graphical}).


\noindent\textbf{(ii)}  We benchmark \fetseg against leading SoTA methods: BOUNTI \citep{uus2022automated}, the FeTA 2024 winner \citep{zalevskyi2025advancesautomatedfetalbrain}, and an nnU-Net ensemble \citep{isensee2021nnu}, on 348 subjects spanning multiple scanners, SR algorithms, and modalities. 

\noindent\textbf{(iii)} We show that \fetseg enables new imaging applications by robustly segmenting T2-weighted images across varying echo times and T1-weighted acquisitions, making new downstream tasks such as T2 mapping in fetal brain imaging possible.

The code, trained models, and a ready-to-use Docker image are available at:
\url{https://github.com/Medical-Image-Analysis-Laboratory/FetalSynthSeg}.

\section{Related Works}
\subsection{Tackling domain shifts}
Various strategies have been proposed to address domain shifts in medical imaging. Data augmentation and fine-tuning are widely used solutions~\citep{Guan2022}, but many task-specific approaches have been designed based on a wide variety of techniques like meta-learning~\citep{Li2018_metalearning}, harmonization~\citep{HU2023120125} or style transfer~\citep{zhou2021domain}. Although effective, these methods often rely on target domain data and annotations, which are costly and difficult to obtain in fetal MRI~\citep{payette2023fetal}. Manual segmentation of a single case can take several hours (e.g., eight hours in~\cite{kyriakopoulou2017normative}). Moreover, approaches requiring multiple source domains are limited when only one domain is available or when existing domains fail to represent new ones~\citep{zhou2022domain}. This limitation is particularly pronounced in fetal imaging, where new scanner field strengths and SR methods introduce previously unseen data distributions. Poor generalization to these domains hinders the adoption of low-field ($<$1T) and ultra-low-field ($<$0.1T) MR systems, which could democratize access to advanced neuroimaging techniques to a broader population~\citep{AvilesVerdera2023}. Furthermore, the scarcity and the small size of current fetal brain MRI datasets exacerbate this problem~\citep{payette2023fetal, Lajous2022}.

\subsection{Single-source domain generalization (SSDG)}
To mitigate these issues, SSDG techniques aim to train models on a \textit{single source domain} while ensuring generalization to unseen target domains. In medical imaging, common strategies include texture and intensity augmentations and synthetic view blending. For example, \citet{ouyang2022causality} proposed a global intensity non-linear augmentation (GIN) module using shallow convolutional networks to enhance intensity diversity, while \citet{Li23_freqmixed_ssdg} simulated inter-domain frequency discrepancies through frequency mixing and self-supervised learning. Other notable efforts include the adversarial domain synthesizer (ADS) by \citet{Xu2022}, which generated synthetic domains using mutual information regularization to preserve semantic consistency, and the test-time augmentation approach by \citet{Liu2022_ssdg_tta}, which leveraged dictionary learning to extract semantic priors for segmentation refinement.

Despite advances, SSDG methods remain limited by the lack of semantic diversity in training data and their focus on 2D images. The FeTA Challenge 2021 clearly showed that 3D models outperform 2D ones for fetal brain segmentation, emphasizing the need for SSDG methods that extend to 3D data~\citep{payette2023fetal}.

\subsection{Synthetic data and domain randomization}
Synthetic data generation and domain randomization have emerged as effective ways to overcome SSDG limitations~\citep{ALKHALIL2023102688, Paproki2024_synthdata1, gopinath2024synthetic}. Deep generative models such as generative adversarial networks (GANs) and diffusion models can produce realistic data but require extensive datasets and are challenging to train~\citep{KAZEROUNI2023102846, KAZEMINIA2020101938, kaandorp2025pathological}. Alternatively, numerical phantoms such as \fabian generate realistic MR images through physical simulations based on anatomical priors without pre-training~\citep{Lajous2022}, but at the cost of longer generation times. \fabian can simulate multiple acquisitions for the same subject with different physical parameters and model fetal motion, allowing to complement scarce clinical datasets~\citep{bhattacharya2024vivo, lajous_2024_datasetzenodo, Lajous2024_maturinform}.

In contrast, \texttt{SynthSeg}~\citep{billot2021synthseg} adopts a domain randomization paradigm: instead of relying on a costly physical modeling, it generates anatomically plausible but visually unrealistic synthetic MRIs with fully randomized contrasts and artifacts~\citep{tobin2017domain} using Gaussian mixture models. The resulting diversity allows to train models that become contrast-agnostic and robust to real-world variability. \texttt{SynthSeg} has been widely used across various MRI tasks, including brain segmentation, skull stripping, and anatomical representation extraction of healthy and pathological subjects~\citep{omidi2024unsupervised, dsynthstrip, pepsi, Liu_2023_BrainID, infantsynthstrip, valabregue2024comprehensive, gopinath2024synthetic}. Importantly, because \texttt{SynthSeg} relies only on anatomical annotations rather than intensity images, the generation process is largely independent of the source imaging domain, aside from anatomical or population differences.

Most synthetic-data-based methods, however, target adult or infant brains. Fetal MRI introduces unique challenges such as a fast-changing morphology across gestation, limited tissue contrast, fewer label classes than in adult imaging, strong motion and artifacts, making realistic simulation and robust segmentation of these data particularly challenging.

\subsection{Fetal brain tissue segmentation}
The FeTA Challenge, organized at MICCAI, has been instrumental in benchmarking fetal brain segmentation algorithms~\citep{payette2024multicenter, zalevskyi2025advancesautomatedfetalbrain}. The 2024 edition emphasized the importance of data augmentation and topology-enhancing post-processing. Top-performing methods used ensembles of 3D architectures such as nnU-Net, and incorporated synthetic or augmented data like \texttt{SynthSeg}~\citep{billot2023synthseg} and GIN~\citep{ouyang2022causality}. The winning team in 2024, \texttt{cesne-digair}, achieved a mean Dice of 0.816 and 95th percentile Hausdorff Distance (HD95) of 2.317, using a 3D U-Net trained on skull-stripped, affine-registered T2w images with hand-crafted augmentations and synthetic registration-based sample generation. A denoising autoencoder was then used to correct segmentation artifacts.

\begin{table*}[t]
\scriptsize
\centering
\caption{\textcolor{black}{Dataset properties. $N_n$ – number of neurotypical subjects, $N_p$ – number of pathological subjects. SR algorithms are MIALSRTK (MIAL~\citep{tourbier_mialsuperresolutiontoolkit_2020}), IRTK~\citep{kuklisova-murgasova_reconstruction_2012}, NiftyMIC~\citep{ebner_automated_2020}, and SVRTK~\citep{uus2022automated}}}
\label{tab:datasets}
\resizebox{1.0\linewidth}{!}{%
\begin{tblr}{
  hline{1-2,8} = {-}{},
  column{5} = {fg=black},
}
 \textbf{Site} & \textbf{Scanner}                                               & {\textbf{Acquisition}\\\textbf{Parameters}}                    & {\textbf{SR}\\\textbf{algorithm}} & {\textbf{Resolution}\\\textbf{($mm^3$)}} & {\textbf{GA }\\\textbf{(weeks)}} & \textbf{$N_n/N_p$} \\
 {KISPI\\ \tiny(public)}        & {GE Signa Discovery \\ MR450/MR750 (1.5T/3T)}            & {SS-FSE \\ TR/TE: 2500–3500/120 ms \\ 0.5 × 0.5 × 3.5 mm³}     & {MIAL \\ IRTK}                    & {$0.5 \times 0.5 \times  0.5$\\$0.5 \times 0.5 \times  0.5$}                                    & {20–34 \\ 20–34}                 & {25/15 \\ 24/16}   \\
 {VIEN\\ \tiny(private)}          & {Philips Ingenia/Intera \\ (1.5T) Philips Achieva (3T)*} & {SS-FSR, TR/TE: \\ 6000–22000/80–140 ms}                               & NiftyMIC                          & $1.0\times 1.0\times1.0$                                               & 19–35                        & 33/7               \\
 {CHUV\\ \tiny(private)}          & {Siemens MAGNETOM \\ Aera (1.5T)}                        & {HASTE, TR/TE: \\ 1200/90 ms \\1.13 × 1.13 × 3 $mm^3$}       & MIAL                              & $1.1 \times 1.1  \times1.1$                                               & 21–35                            & 25/15              \\
 {KCL\\ \tiny(private)}           & {Siemens MAGNETOM \\ FREE.MAX (0.55T)}                   & {HASTE, TR/TE: \\ 2600/106 ms \\1.45 × 1.45 × 4.5 $mm^3$}    & {SVRTK \\}                 & $0.8 \times 0.8 \times 0.8$                                               & 21–35                            & 15/5               \\
 {dHCP (T2w)\\ \tiny(public)}    & {Philips Achieva (3T)}                                   & {MB-TSE, TR/TE: \\ 2265/250 ms \\1.1 × 1.1 × 2.2 $mm^3$}     & IRTK                              & $0.8 \times 0.8 \times 0.8$                                              & 21–38                            & 248/0              \\
 {dHCP (T1w)\\ \tiny(public)}    & {Philips Achieva (3T)}                                   & {bSSFP, TR/TE: \\ 3.6/7.2 (8479*) ms \\1.5 × 1.5 × 4 $mm^3$} & IRTK                              & $0.8 \times 0.8 \times 0.8$                                               & 21–38                            & 208/0              
\end{tblr}
}
\vspace{-.3cm}
\begin{flushleft}
\footnotesize
\hspace{1cm} *Slice package duration (effective TR for slab selective IR pulse)
\end{flushleft}
\end{table*}

Using FeTA’s public dataset, \citet{s23020655} proposed a hybrid convolution-transformer achieving $0.837\pm0.03$ Dice on a held-out \textit{in-domain} validation set.  Beyond the FeTA dataset, BOUNTI~\citep{Uus2023.04.18.537347_Bounti} combined U-Net and Attention U-Net trained on 380 fetal MRIs (public and private), reaching $0.89\pm0.02$ Dice, thanks to extensive manual refinement of ground truth segmentations that contributed to the model’s performance. However, its reliance on private datasets and custom annotation protocols limits its comparability with FeTA-trained models.

These studies show that high performance can be achieved through extensive data augmentation, the inclusion of diverse multi-site data, and expert-curated labels. Yet, such resources are not always available. In contrast, our work investigates a more common and resource-constrained scenario: \textbf{how to train fetal brain segmentation models that generalize well when using only public data from a single-domain with standard annotation quality}. We aim to characterize the limitations of this setting and propose practical strategies that enhance generalization in data-scarce environments.

\section{Methods}

\subsection{Data}
\label{data}

We use open-access data from the FeTA challenge~\citep{payette2023fetal} and the developing Human Connectome Project (dHCP)~\citep{Edwards2022}, complemented with private clinical datasets from three institutions (Table~\ref{tab:datasets}). All acquisitions were approved by local ethics committees and performed without maternal or fetal sedation.
For all datasets, several 2D single-shot fast spin-echo stacks were acquired in at least three orthogonal orientations (axial, coronal, sagittal) to mitigate motion artifacts (Table~\ref{tab:datasets}, \textit{Acquisitions parameters}). A super-resolution reconstruction algorithm then combined these stacks into a high-resolution 3D volume suitable for volumetric segmentation.

\paragraph{FeTA Challenge Datasets.}
We use subsets of both training and testing datasets employed in the FeTA 2024 Challenge. Acquisition protocols and site-specific characteristics for KISPI, VIEN, CHUV, and KCL are detailed in Table~\ref{tab:datasets} and in~\citet{zalevskyi2024improving, payette2023fetal}. \textcolor{black}{Among these, only the KISPI dataset is publicly available on Synapse \citep{FeTA_Synapse}. In contrast, CHUV and KCL were used as held-out test datasets in the FeTA challenge and were not shared with participants, serving exclusively for final evaluation. Although VIEN was part of the challenge training set, its use is restricted to the scope of the challenge, and it is therefore not publicly available. Consequently, we refer to CHUV, KCL, and VIEN as private datasets.}

\paragraph{Fetal dHCP Dataset.}
The fetal dHCP dataset, acquired at St Thomas’ Hospital, London, includes T2w and T1w structural images obtained on a Philips Achieva 3T scanner with a 32-channel cardiac coil~\citep{price2019developing, karolis2025developing}. 
T1w data were acquired using eight stacks across six unique orientations with an inversion-recovery sequence and interleaved wide-slab preparation pulses, followed by a bSSFP readout (in-plane resolution 1.5$\times$1.5 mm$^2$, slice thickness 4 mm, slice gap $-1.2$ mm, flip angle 35°, total scan $\sim$8 min). T2w data were acquired using a zoomed multiband single-shot TSE sequence (in-plane resolution 1.1$\times$1.1 mm$^2$, slice thickness 2.2 mm, slice gap $-1.1$ mm, flip angle 30° or 130°, total scan $\sim$12 min). All images were reconstructed to 0.5 mm isotropic using the fetal branch of the dHCP structural pipeline~\citep{Makropoulos125526dhcprecon, cordero2022fetal}.

\paragraph{Prospective multi-echo KCL dataset}
Data were prospectively acquired for this study from five fetuses (GA = 24, 32, 32, 33, 37 weeks) using SST2w sequences at St. Thomas' Hospital, London, on a Siemens FreeMax 0.55T scanner, with similar parameters as the KCL dataset in Table~\ref{tab:datasets}, except for echo times (TEs), with data acquired at 300 ms, 397 ms, 600 ms, in order to do T2 mapping. At each echo time, three stacks (axial, coronal, sagittal) were acquired. The study was conducted under the ethically approved MEERKAT [REC: 21/LO/0742], MiBirth [REC: 23/LO/0685], and NANO [REC: 22/YH/0210] projects. Data sharing was approved by the Ethics Committee London Bromley (Ethics code 21/LO/0742). The data were then reconstructed using the method of \citet{bulut2025physics}, which yielded a volume at each TE.

\paragraph{Pre-processing and labeling.}
All structural images used for training and inference were resampled to 0.5 mm isotropic resolution and standardized to a size of $256^3$ voxels via cropping or zero-padding.
For FeTA datasets, we used the original FeTA labels comprising seven classes: cerebrospinal fluid (CSF), white matter (WM), gray matter (GM), subcortical gray matter (SGM), ventricles (LV), brainstem (BSM), and cerebellum (CBM)~\citep{payette_automatic_2021}. For dHCP, we remapped Draw-EM annotations~\citep{MAKROPOULOS201888, Makropoulos125526dhcprecon} to the FeTA scheme using:
1$\rightarrow$CSF, 2$\rightarrow$GM, 3$\rightarrow$WM, 4$\rightarrow$background, 5$\rightarrow$LV, 6$\rightarrow$CBM, 7$\rightarrow$SGM, 8$\rightarrow$BSM,\linebreak 9$\rightarrow$WM.
Thus, hippocampi and amygdala (9) are merged with WM, and the skull (4) with the background. Draw-EM and FeTA differ slightly in anatomical definitions (e.g., ventricles coverage), but all final datasets share a unified seven-tissue labeling consistent with FeTA.

\subsection{Synthetic Data Generation Frameworks}
We start by testing Hypothesis~\ref{hyp:h1} by comparing various synthetic data generation frameworks, to assess how physics-based and domain randomization-based frameworks behave as data augmentation strategies. Our experimental framework builds on three components: \synthseg~\citep{billot2021synthseg}, the \fabian numerical phantom~\citep{Lajous2022}, and \fetseg, our fetal-tailored extension of \synthseg~\citep{zalevskyi2024improving}. These components isolate how different intensity generation strategies impact single-source domain generalization. Note that these components can generate images that appear unrealistic, as illustrated in Figure~\ref{fig:graphical}. Exaggerated anatomies or non-physical contrasts, artifacts, and resolutions are an intentional feature of domain randomization, and the value of these generated images will be assessed through the downstream segmentation performance of models using them throughout training, not through their visual appearance. Overall, for each synthetic data generation approach, we follow a two-stage process: (1) generate synthetic images using different intensity simulation strategies, and (2) train a standard 3D U-Net segmentation network on these images.

\begin{figure*}[t]
    \centering
    \includegraphics[width=1\linewidth]{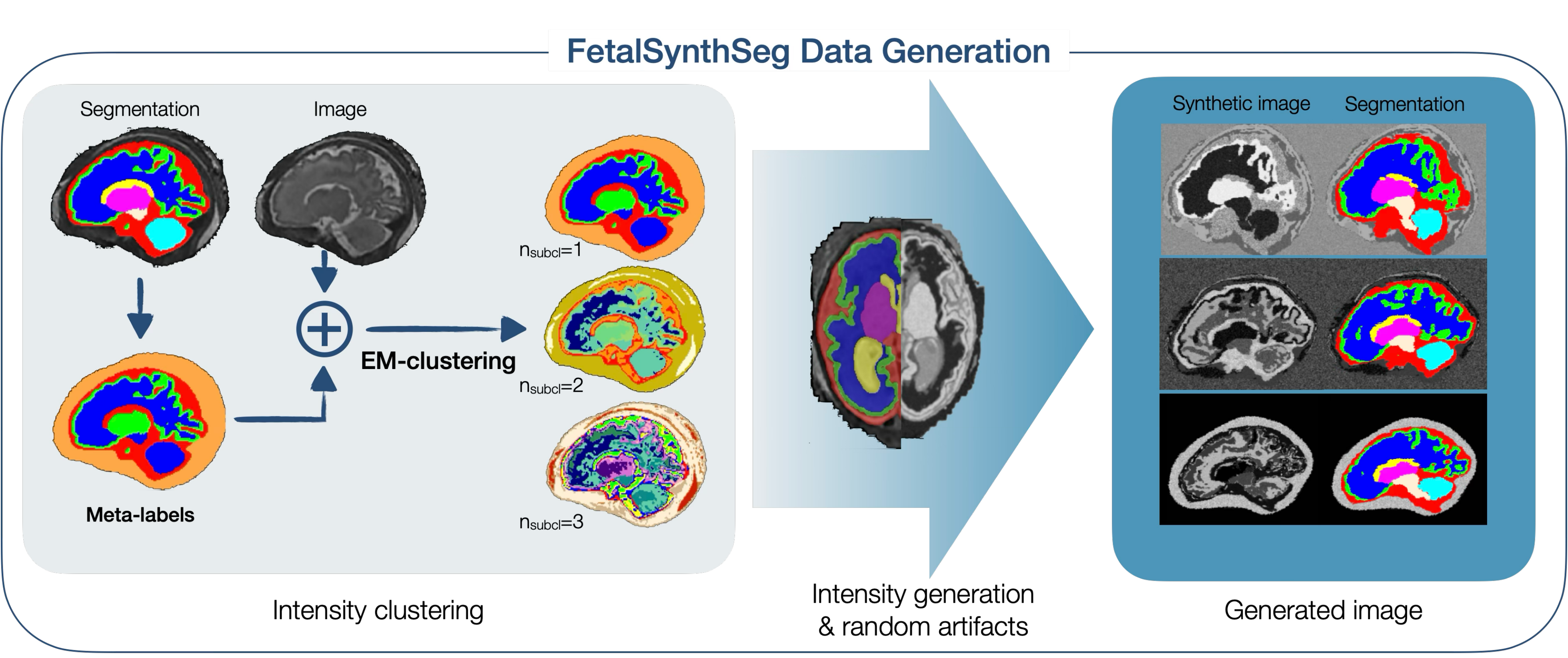}
    \caption{\textbf{\fetseg's pipeline for meta-label splitting into sub-classes used for training image generation.} Original segmentation labels are merged into four meta-labels (WM, GM, CSF, non-brain tissue). Then, each meta-label is randomly split into sub-classes based on EM-clustering using the original image. Finally, an independent GMM is sampled for each of the sub-classes and used to sample intensities for voxels inside of it. Synthesized training images are then fed into the U-Net. During testing only real images are used for inference.}
    \label{fig:meta-sub_classes}
\end{figure*}

\subsubsection{SynthSeg}
\label{synthseg}

\texttt{SynthSeg} generates synthetic MR images directly from label maps using domain randomization~\citep{billot2021synthseg}. Given segmentations $\{S_n\}$, a segmentation is randomly sampled, spatially transformed by $\phi$ (affine and non-linear diffeomorphic transforms), and intensities at voxel $(x,y,z)$ are drawn from a Gaussian mixture model (GMM) with label-specific parameters:
\[
G(x, y, z) \sim \mathcal{N}\big(\mu_{L(x, y, z)}, \sigma_{L(x, y, z)}^2\big),
\]
where $\mu_L \sim \mathcal{U}(a_\mu,b_\mu)$ and $\sigma_L \sim \mathcal{U}(a_\sigma,b_\sigma)$.  The image is then corrupted by adding a bias field, intensity transformations and by simulating various image resolutions.  By sampling a broad range of appearances beyond what is realistic, \synthseg promotes robustness to the severe domain shifts faced in practice.

\subsubsection{FetalSynthSeg}
\label{fss}

FeTA annotations contain only seven tissue classes, which is coarse compared to adult segmentations with 30 classes used in \synthseg. This limits the ability to reproduce heterogeneous fetal tissue appearances, particularly within developing WM~\citep{Lajous2024_maturinform}. \fetseg addresses this by augmenting the label space via unsupervised intensity-based clustering.
Following \citet{zalevskyi2024improving}, we first group labels into four meta-labels (or meta-classes):
\textbf{WM} (WM, CBM, BSM),
\textbf{GM} (cortical and deep GM),
\textbf{CSF} (ventricles and external CSF),
\textbf{non-brain} (skull, uterus, fetal body, maternal tissue). This grouping fuses labels for structures that have similar macro-structural composition and hence have similar intensity profiles on the structural images during the late gestation period.
All non-zero voxels are assigned to one of these meta-labels, allowing realistic modeling of both brain and surrounding anatomy as encountered in fetal pipelines (Figure~\ref{fig:meta-sub_classes}).

Each meta-label is then \textbf{randomly and independently partitioned} into a number of subclasses sampled from $\mathcal{U}(a_{\text{subcl}}, b_{\text{subcl}})$. Sub-clustering is performed with an Expectation–Maximization algorithm conditioned on the input intensities~\citep{em_algo} (Fig.~\ref{fig:meta-sub_classes}). Each subclass is assigned its own GMM parameters, introducing controlled intra-class variability. Importantly, intensity-based clustering itself is not new and has previously been used in domain-randomisation approaches such as SynthSeg \citep{billot2021synthseg}. In contrast to prior work that performs clustering within existing anatomical labels, our approach performs clustering within meta-labels that group anatomically related tissues and background structures. This allows subclasses to span boundaries between structures with similar intensity profiles (e.g., WM and CBM) while preserving realistic contrast transitions. In the fetal setting, where annotations are coarse and tissue appearance is highly heterogeneous, this strategy enables more flexible modelling of intensity variability without introducing artificial boundaries.

\paragraph{Implementation.}
\synthseg and \fetseg are implemented using the PyTorch SynthSeg generator from Brain-ID~\citep{Liu_2023_BrainID}, with hyperparameters kept identical to the original implementation. Our generator and modifications are available at:
\url{https://github.com/Medical-Image-Analysis-Laboratory/fetalsyngen}.

\subsubsection{\fabian and \randfabian}

\fabian is an open-source numerical phantom that simulates clinical T2w FSE fetal brain images using tissue segmentations and the extended phase graph (EPG) formalism~\citep{Lajous2022}. It models signal evolution based on literature T1/T2 values and sequence parameters, generating WM, GM, and CSF contrasts based on the physical tissue decay properties.

We adapt \fabian to include the non-brain meta-class consistent with \fetseg, and draw T1/T2 values from reference distributions before physical simulation, introducing controlled variability. We further define \randfabian, where all T1/T2 values are sampled from a broad, unconstrained range, making it more strongly randomized and conceptually closer to \fetseg.

\paragraph{Implementation.}
We use the Matlab implementation of \texttt{FaBiAN\_v2}~\citep{Lajous2024_maturinform, lajous_2024_datasetzenodo} via a Python wrapper:
\url{https://github.com/Medical-Image-Analysis-Laboratory/fabian_utils}.
Examples for all generators are shown in Fig.~\ref{fig:graphical}, and details of the parameters used are available in the Supplementary Table~\ref{table:haste_params}.

\subsection{Baselines: Real Image Training and Hybrid Training}

After evaluating the data generation process, we move onto Hypothesis~\ref{hyp:h2}, where we aim to benchmark the ID and OOD performance of domain-randomization-based models. To quantify the value of randomized intensity simulation over conventional model training, we train two baseline models: 
\begin{enumerate}
    \item \frs is trained exclusively on real images using the same augmentation pipeline as synthetic-data models. \frs uses precisely the real scans whose segmentations serve as anatomical priors for the synthetic generators. Apart from the absence of randomized intensities, all transformations and training settings are identical, enabling a controlled comparison between purely real-data training and synthetic-data-driven approaches.
    \item \textcolor{black}{\hrs is trained using a mixture of 50$\%$ synthetic and 50$\%$ real data, and is essentially a hybrid between \fetseg and \frs, randomly generating either a random synthetic sample or a data-augmented intensity image, each with 50$\%$ probability.}
\end{enumerate}

\subsection{Experimental Design}
We conduct \textcolor{black}{four} main experiments:

\paragraph{I. Selecting an intensity generation strategy (\ref{exp:1})}
We benchmark \frs, \fabian, \randfabian, \synthseg, and \fetseg in a single-source domain generalization setting. Due to the computational cost of \fabian, this experiment uses three domains: KISPI-IRTK (n=40), KISPI-MIAL (n=40), and CHUV-MIAL (n=40). We train on one domain and test on the remaining two, covering mild (site or SR-only) and severe (site+SR) domain shifts. The goal is to identify which generator yields the most robust out-of-domain performance.

\paragraph{II. Impact of intensity clustering (\ref{exp:2}).}
We perform an ablation on the impact of \textit{meta-label fusion and EM-clustering} by varying the number of EM clusters in \synthseg and \fetseg. Using the splits from Experiment~I, we assess how sub-clustering impacts performance and whether the \fetseg design consistently improves robustness.

\paragraph{III. Comparison with state-of-the-art models with failure mode analysis (\ref{exp:3}).}
We retrain \fetseg, \textcolor{black}{\frs and \hrs} on the full FeTA 2024 training set (n=120) and compare them against:
\begin{enumerate}[itemsep=1pt, parsep=0pt,leftmargin=0.8cm]
    \item the FeTA 2024 winning method \texttt{cesne-digair}\linebreak  (\textit{FeTA24})~\citep{zalevskyi2025advancesautomatedfetalbrain}
    \item BOUNTI~\citep{Uus2023.04.18.537347_Bounti},
    \item an nnU-Net ensemble~\citep{isensee2021nnu}.
\end{enumerate}
 All models are evaluated on 348 subjects from the dHCP dataset and a subset of FeTA 2024 test data, spanning T1w/T2w, 0.55T/1.5T/3T, and both pathological and neurotypical cases.  \textcolor{black}{We use the default implementations and publicly available weights for the FeTA 2024 challenge winner\footnote{https://hub.docker.com/u/fetachallenge2024} and the nnU-Net model \footnote{https://github.com/mic-dkfz/nnunet}, both trained on the full FeTA 2024 training dataset (n = 120). For BOUNTI, we use the publicly available Docker image \footnote{https://hub.docker.com/r/fetalsvrtk/segmentation} with weights trained on a larger collection of private and public datasets unrelated to the FeTA 2024 Challenge (n = 360).}
 
\paragraph{IV. Prospective validation (\ref{exp:4}).} \textcolor{black}{Finally, we showcase the abilities of \fetseg to tackle challenging data by validating its ability to consistently segment prospectively acquired data at different echo times (TEs)~\citep{bulut2025physics}.}

\subsection{Model Architecture and Training}

All models use the same 3D U-Net backbone, implemented in PyTorch with MONAI~\citep{paszke2019pytorch, cardoso2022monai}. The network has five levels with 32 feature maps at the first level (doubling per level), $3^3$ convolutions with LeakyReLU activations, skip connections, and a softmax output layer. We deliberately use a simple, consistent architecture to focus the analysis on data generation strategies rather than architectural optimizations.

We train with Adam (learning rate $10^{-3}$) and a combined Dice + cross-entropy loss~\citep{valabregue2024comprehensive}. A \texttt{ReduceLROnPlateau} scheduler (factor 0.1, patience 10 epochs) and \texttt{EarlyStopping} (patience 100 iterations) are applied. Batch size is 1. Training is implemented with PyTorch Lightning and run on NVIDIA RTX 3090/6000 GPUs.

For \synthseg and \fetseg, synthetic images are generated on-the-fly for up to 80,000 iterations. For \fabian and \randfabian, we pre-generate 50 synthetic images per real scan (n=120; 6000 images total) and train for up to 80,000 iterations. Online generation with \fabian is infeasible ($\sim$287 s per volume) compared to $\sim$1 s for \synthseg/\fetseg. \textcolor{black}{For \frs and \hrs, we train on for up to 500 epochs to limit overfitting.}  

\begin{table*}[t]
\centering
\caption{Mean Dice scores (multiplied by 100 for readability) for models trained on different data sources, evaluated across multiple testing splits. Each cell reports the mean and standard deviation across test subjects. \textbf{Bold} indicates the best-performing method per column, and \underline{underlined} denotes the second-best. An asterisk ($^*$) marks cases where the best method is statistically significantly better than the second best according to a Wilcoxon rank-sum test ($p<0.05$).}

\label{tab:2_results_fssVSALL}
\resizebox{.9\linewidth}{!}{%
\begin{tblr}{
  column{even} = {c},
  column{3} = {c},
  column{5} = {c},
  column{7} = {c},
  cell{1}{2} = {c=2}{},
  cell{1}{4} = {c=2}{},
  cell{1}{6} = {c=2}{},
  cell{1}{8} = {r=2}{},
  hline{1,8} = {-}{0.08em},
  hline{2} = {1-3}{},
  hline{2} = {4-7}{l},
  hline{3} = {-}{0.05em},
  hline{7} = {-}{0.08em},
}
\textbf{Testing split}  & \textbf{CHUV-MIAL}    &                       & \textbf{KISPI-IRTK}   &                       & \textbf{KISPI-MIAL} &                      & Global                 \\
\textbf{Training split} & KISPI-IRTK & KISPI-MIAL & CHUV-MIAL & KISPI-MIAL & CHUV-MIAL & KISPI-IRTK & \\
\fabian & 74.2{\scriptsize ±4.2} & 73.1{\scriptsize ±5.7} & 53.6{\scriptsize ±13.3} & 56.6{\scriptsize ±12.9} & 60.6{\scriptsize ±17.1} & 61.5{\scriptsize ±20.1} & 63.3{\scriptsize ±15.5} \\
\randfabian & \uline{79.1{\scriptsize ±2.3}} & \uline{78.2{\scriptsize ±2.9}} & 55.1{\scriptsize ±12.7} & 68.9{\scriptsize ±7.8} & 60.7{\scriptsize ±7.3} & \uline{68.1{\scriptsize ±14.7}} & 68.3{\scriptsize ±13.8} \\
\synthseg & 75.9{\scriptsize ±3.9} & 73.7{\scriptsize ±3.9} & \uline{70.9{\scriptsize ±9.2}} & \uline{74.8{\scriptsize ±7.8}} & 60.5{\scriptsize ±15.7} & 63.4{\scriptsize ±16.8} & 72.2{\scriptsize ±13.0} \\
\fetseg & $\textbf{80.7{\scriptsize ±2.0}}^*$ & 76.9{\scriptsize ±3.3} & $\textbf{79.2{\scriptsize ±9.0}}^*$ & $\textbf{76.8{\scriptsize ±6.9}}^*$ & \uline{67.5{\scriptsize ±16.0}} & $\textbf{68.5{\scriptsize ±15.6}}$ & $\textbf{74.9{\scriptsize ±11.5}}^*$ \\
\frs & 77.2{\scriptsize ±4.0} & $\textbf{78.5{\scriptsize ±3.3}}^*$ & 70.6{\scriptsize ±13.9} & 71.9{\scriptsize ±11.5} & \textbf{68.0{\scriptsize ±19.3}} & 64.3{\scriptsize ±19.1} & \uline{73.0{\scriptsize ±12.6}} \\
    
\end{tblr}
}

\end{table*}
\begin{figure*}[h]
    \vspace{-.5cm}
    \centering
    \includegraphics[width=1\linewidth]{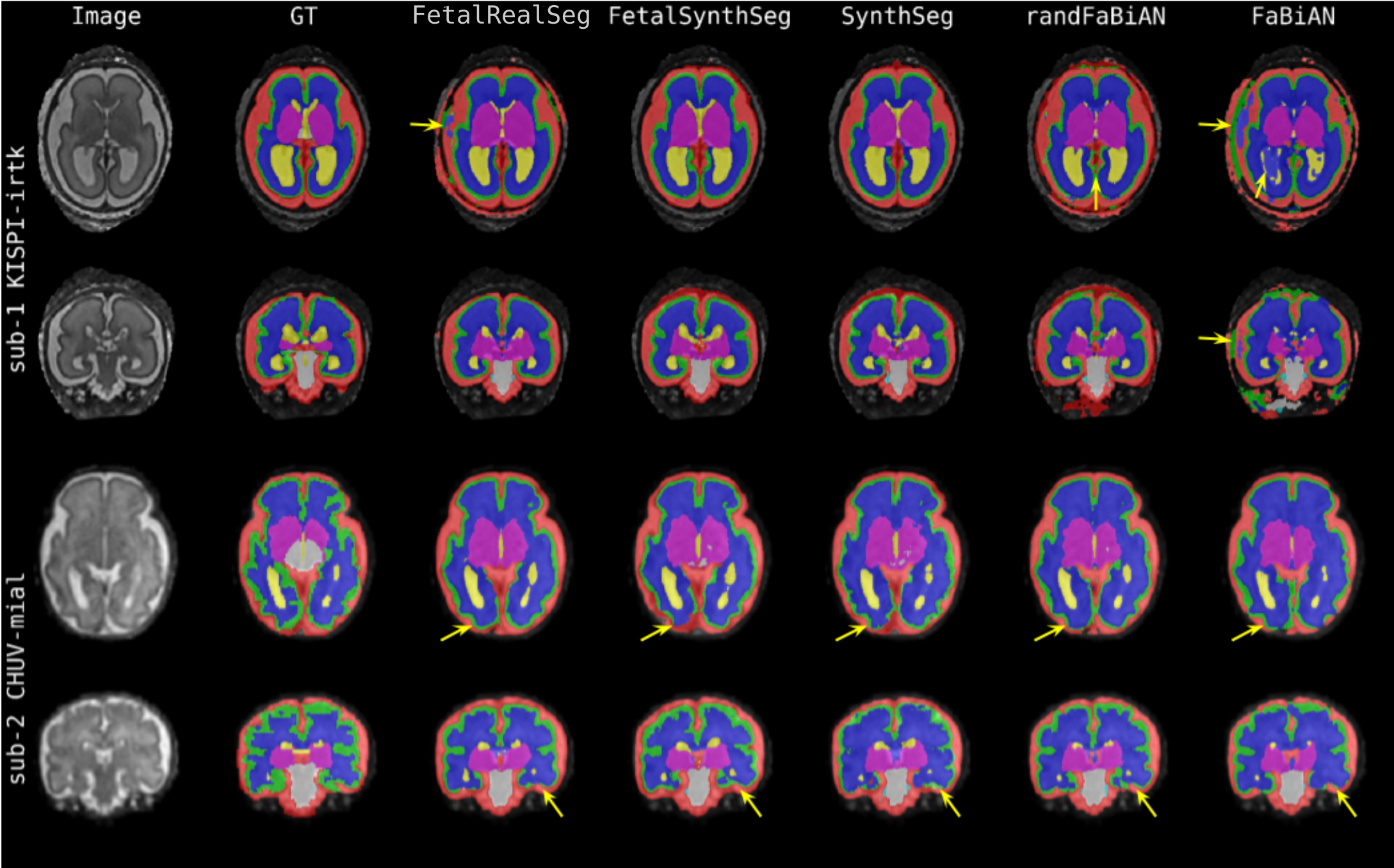}
    \vspace{-.5cm}
    \captionof{figure}{Segmentation results for models trained on the KISPI-MIAL data split, evaluated on a subject from the KISPI-IRTK split and on another from the CHUV-MIAL split. Arrows point to regions of substantial discrepancies across models.}
    \vspace{-.5cm}
    \label{fig:qualitative_feta}
\end{figure*}

\paragraph{Data augmentation.}
All models use the \synthseg augmentation pipeline: random non-linear deformations, gamma-based contrast changes, Gaussian noise, bias field perturbations, and random isotropic resampling to emulate varying resolutions.

\subsection{Evaluation Protocol}

\paragraph{Data splitting.}
For Experiment~1, we perform cross-domain evaluation by training on one of KISPI-IRTK, KISPI-MIAL, or CHUV-MIAL (40 scans each) and testing on the other two domains. For each training domain, 35 scans are used for training and 5 for validation; all 40 scans of each held-out domain are used for testing. We report per-domain and pooled performance.  Experiment~2 reports results broken down per split, while Experiment~3 relies on standard train/validation/test split.

\paragraph{Metrics.}
We evaluate segmentations using:
\begin{itemize}
  \item \textbf{Dice Similarity Coefficient (DSC):}
  \[
  DSC = \frac{2|A \cap B|}{|A| + |B|},
  \]
  with $A$ and $B$ the predicted and ground-truth voxel sets.
  \item \textbf{95th-percentile Hausdorff Distance (HD95):}
  \[
  HD95 = \max\left(
    \max_{x \in A} \min_{y \in B} \|x-y\|,\,
    \max_{y \in B} \min_{x \in A} \|x-y\|
  \right),
  \]
  with $A$ and $B$ boundary point sets. 
\end{itemize}

\textcolor{black}{When not stated otherwise, the metrics that we report are first averaged across the 7 labels for each subject, and then averaged across all subjects in experiments where we compare different models.}

\paragraph{Statistical analysis.}
Normality is assessed with the\linebreak Shapiro–Wilk test. Depending on distribution, we use either a paired \mbox{t-test} or Wilcoxon rank-sum test with Bonferroni correction. Results with $p < 0.05$ are considered statistically significant.

\section{Results}

\subsection{Selecting an intensity generation strategy}\label{exp:1}
A global comparison of all intensity generation methods is reported in Table~\ref{tab:2_results_fssVSALL}. Across all cross-validation splits, \fetseg achieves the highest global average Dice score of 74.9, consistently outperforming all other simulation-based approaches. \textcolor{black}{Across individual splits, the only significant exception is the KISPI-mial$\to$CHUV-mial split, where \frs outperforms \fetseg, suggesting that training on real images can provide advantages under certain domain shift conditions (same SRR method in this experiment). Overall, \frs remains the second-best performing model, with an average Dice of 73.0, indicating that synthetic data generation alone does not necessarily guarantee improved generalization, and models trained on real images may remain competitive when the domain gap between training and testing data is relatively limited.}

\begin{figure}
    \centering
    \vspace{-.5cm}
    \includegraphics[width=\linewidth]{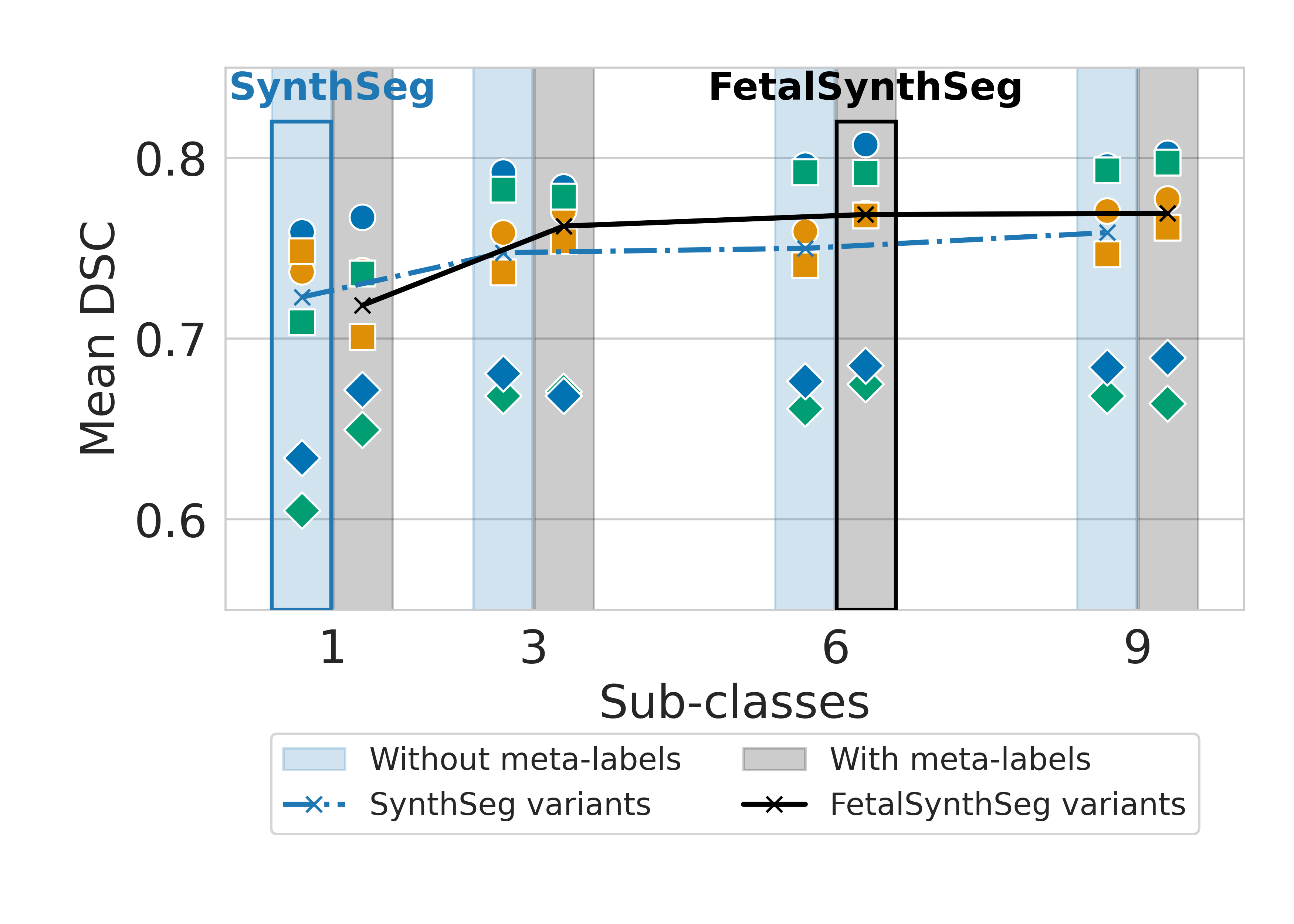}
    \vspace{-1cm}
    
    \caption{\textbf{Ablation study on the number of EM sub-clusters.} For \texttt{SynthSeg}-variants, the original 7 labels used as generation classes are randomly split into 1-9 subclasses, while in \texttt{FetalSynthSeg}-variants the meta-labels are split. \textcolor{black}{The blue and black rectangles show respectively what we refer to as \synthseg (one sub-class with no meta-labels) and \fetseg (six sub-classes with meta-labels) in the paper.}  We report mean Dice across all tissues for each combination of training-testing splits. The different labels denote these splits:\linebreak
    \resizebox{\linewidth}{!}{
    \begin{tabular}{ll}
        {\color{myorange}\ding{108}} CHUV-MIAL$\rightarrow$KISPI-IRTK&{\color{myblue}\ding{108}}
        CHUV-MIAL$\rightarrow$KISPI-MIAL\\
        {\color{myorange}\ding{110}} KISPI-MIAL$\rightarrow$KISPI-IRTK& 
        {\color{mygreen}\ding{110}} KISPI-MIAL$\rightarrow$CHUV-MIAL\\
        {\color{myblue}\ding{117}}~KISPI-IRTK$\rightarrow$KISPI-MIAL&
        {\color{mygreen}\ding{117}} KISPI-IRTK$\rightarrow$CHUV-MIAL 
    \end{tabular}}
    \vspace{-.5cm}
}
\label{fig:fig4_fss_ss_ablations}
\end{figure}

Focusing on synthetic generation methods, the progression across generation strategies highlights the impact of increasing intensity variability. \fabian, which uses tissue-specific relaxometry values with realistic physics-based modeling, yields the lowest mean Dice (63.3). Randomizing T1/T2 within \randfabian improves performance by +5.0 (68.3). Replacing physics-based simulation with GMM-based sampling in \synthseg leads to a further increase of +3.9 (72.2). Incorporating meta-classes and intensity sub-clustering in \fetseg brings an additional +2.7 gain, reaching a mean Dice of 74.9.

These differences are not solely explained by training set size. Supplementary Experiment~\ref{sec:fabian} shows that, even when matched to \fabian with 6{,}000 synthetic samples, \fetseg still improves by 3.9 Dice points. Moreover, \fabian’s physics-based simulation is computationally demanding ($\sim$280s per volume), whereas the GMM-based sampling of \synthseg and \fetseg requires $\sim$1s, enabling efficient on-the-fly generation during training.

Qualitative results in Fig.~\ref{fig:qualitative_feta} corroborate the quantitative findings: methods with stronger contrast randomization (\randfabian, \fetseg) are noticeably more robust to changes in reconstruction and skull-stripping quality than approaches relying solely on realistic or real intensities.

Per-label Dice and HD95 scores are provided in Supplementary Tables~\ref{tab:2fetalsynthseg_comparative_full} and~\ref{tab:hdtable}. While HD95 differences between \synthseg and \fetseg are small (mean HD95 $3.15 \pm 0.52$ vs. $3.16 \pm 0.52$), all domain-randomized models outperform purely real-data or strictly physics-based approaches, confirming that randomized synthetic intensities are key for robust, cross-domain fetal brain segmentation.

\subsection{Impact of intensity clustering}\label{exp:2}

Figure~\ref{fig:fig4_fss_ss_ablations} summarizes the effect of intensity sub-clustering on \synthseg and \fetseg. Intensity-clustering is done on the seven tissue classes for \synthseg and on the four meta-labels for \fetseg. Starting from the baseline configuration (a single class per label), gradually increasing the number of subclasses consistently improves Dice scores for both methods. Performance gains saturate beyond approximately six subclasses, suggesting an optimal range of sub-cluster granularity.

Across most configurations, \fetseg maintains a systematic advantage over \synthseg, demonstrating that meta-label fusion and sub-clustering act synergistically. Over the four tested sub-cluster settings, \fetseg outperforms \synthseg in 7/12 train–test splits (p-value $<$ 0.05), \synthseg is superior in 3/12, and the remaining 2/12 show no significant difference. This indicates that enriching the label space with structured yet randomized subclasses is a key ingredient for strong SSDG performance.

\textcolor{black}{Furthermore, based on our experiments, performance improvements from additional subclass splitting plateau at six subclasses, so we use this setting for all following experiments involving \fetseg.}

\begin{table*}[t]
    \centering

    \caption{Dice performance (in \%) and surface distance (in mm) across datasets for each model. Values are mean{\scriptsize$\pm$} standard deviation. Bold indicates best and underlined second-best performance per column. \textcolor{black}{An asterisk ($^*$) marks cases where the best method is statistically significantly better than the second-best method according to a Wilcoxon rank-sum test ($p < 0.05$).}}
    \label{tab:diceresSoTA}
\resizebox{\linewidth}{!}{
\begin{tabular}{p{3.3cm}cccccccccc}
\toprule
\multirow{2}{*}{Model} & \multicolumn{5}{c}{\textbf{DSC}} & \multicolumn{5}{c}{\textbf{HD95}} \\
\cmidrule(rl){2-6}\cmidrule(rl){7-11}

 & CHUV$_{T2w}$ & KCL$_{T2w}$ & VIEN$_{T2w}$ & dHCP$_{T2w}$ & dHCP$_{\textbf{T1w}}$ & CHUV$_{T2w}$ & KCL$_{T2w}$ & VIEN$_{T2w}$ & dHCP$_{T2w}$ & dHCP$_{\textbf{T1w}}$ \\
\midrule
BOUNTI   & 79.9{\scriptsize$\pm$2.2} & 83.4{\scriptsize$\pm$1.1} & 70.1{\scriptsize$\pm$12.3} & \textbf{85.1{\scriptsize$\pm$1.3*}} & 16.4{\scriptsize$\pm$2.1} & 3.15{\scriptsize$\pm$0.46} & 2.62{\scriptsize$\pm$0.27} & 4.51{\scriptsize$\pm$2.58} & \textbf{3.56{\scriptsize$\pm$0.42*}} & 15.86{\scriptsize$\pm$3.36}\\
FeTA24   & 83.0{\scriptsize$\pm$1.8} & 85.6{\scriptsize$\pm$0.9} & 81.4{\scriptsize$\pm$4.1} & 82.4{\scriptsize$\pm$1.5} & 23.3{\scriptsize$\pm$3.8} & 2.49{\scriptsize$\pm$0.42} & 1.82{\scriptsize$\pm$0.27} & 2.47{\scriptsize$\pm$0.86} & \uline{4.07{\scriptsize$\pm$0.58}} & 10.35{\scriptsize$\pm$2.03} \\
nnU-Net  & 81.8{\scriptsize$\pm$1.8} & \textbf{86.7{\scriptsize$\pm$1.2}} & \textbf{83.6{\scriptsize$\pm$3.4*}} & \uline{84.3{\scriptsize$\pm$1.2}} & 13.5{\scriptsize$\pm$2.5} & 2.55{\scriptsize$\pm$0.35} & \uline{1.64{\scriptsize$\pm$0.15}} & \textbf{2.07{\scriptsize$\pm$0.54}} & 4.08{\scriptsize$\pm$0.47} & 25.16{\scriptsize$\pm$8.74} \\
\frs      & \textbf{84.6{\scriptsize$\pm$1.8}} & \uline{86.5{\scriptsize$\pm$1.1}} & \uline{82.0{\scriptsize$\pm$3.3}} & 84.1{\scriptsize$\pm$1.5} & 43.7{\scriptsize$\pm$7.0}  & \textbf{1.79{\scriptsize$\pm$0.23}} & \textbf{1.55{\scriptsize$\pm$0.16*}} & \uline{2.11{\scriptsize$\pm$0.50}} & 4.11{\scriptsize$\pm$0.65} & {9.10{\scriptsize$\pm$4.08}}\\
\hrs & \uline{84.2{\scriptsize$\pm$1.5}} & 85.9{\scriptsize$\pm$1.2} & 80.1{\scriptsize$\pm$4.3} &83.1{\scriptsize$\pm$1.7} & \uline{78.1{\scriptsize$\pm$2.5}} & \uline{1.84{\scriptsize$\pm$0.22}} & 1.76{\scriptsize$\pm$0.26} & 2.46{\scriptsize$\pm$0.90} & 4.32{\scriptsize$\pm$0.55} & \uline{4.42{\scriptsize$\pm$0.85}}   \\
\fetseg      & 83.2{\scriptsize$\pm$1.7} & 85.3{\scriptsize$\pm$1.2} & 79.5{\scriptsize$\pm$4.1} & 84.0{\scriptsize$\pm$2.1} & \textbf{80.1{\scriptsize$\pm$2.1*}} & {1.89{\scriptsize$\pm$0.22}} & 1.76{\scriptsize$\pm$0.25} & 2.17{\scriptsize$\pm$0.40} & 4.08{\scriptsize$\pm$0.50} & \textbf{4.27{\scriptsize$\pm$0.53*}}\\
\bottomrule
\end{tabular}}

\end{table*}

\subsection{Comparison with state-of-the-art models}\label{exp:3}

Table~\ref{tab:diceresSoTA} compares \fetseg, \frs and \textcolor{black}{\hrs} (trained on the full FeTA 2024 training set) against BOUNTI, the FeTA24 winning method (\texttt{cesne-digair}), and an nnU-Net ensemble. On T2w datasets that are close to the training distribution (e.g., CHUV$_{T2w}$; VIEN$_{T2w}$ sharing both scanner/SRR; KCL with SVRTK being similar to IRTK), \frs and \textcolor{black}{\hrs} remain slightly ahead of \fetseg, with differences typically within 0.1–2.5 Dice points. \fetseg, \textcolor{black}{\hrs} and \frs are competitive with or superior to existing SoTA models in these settings: for instance, on CHUV, they surpass all reference methods.

\begin{figure*}[ht!]
    \centering
    \includegraphics[width=1\linewidth]{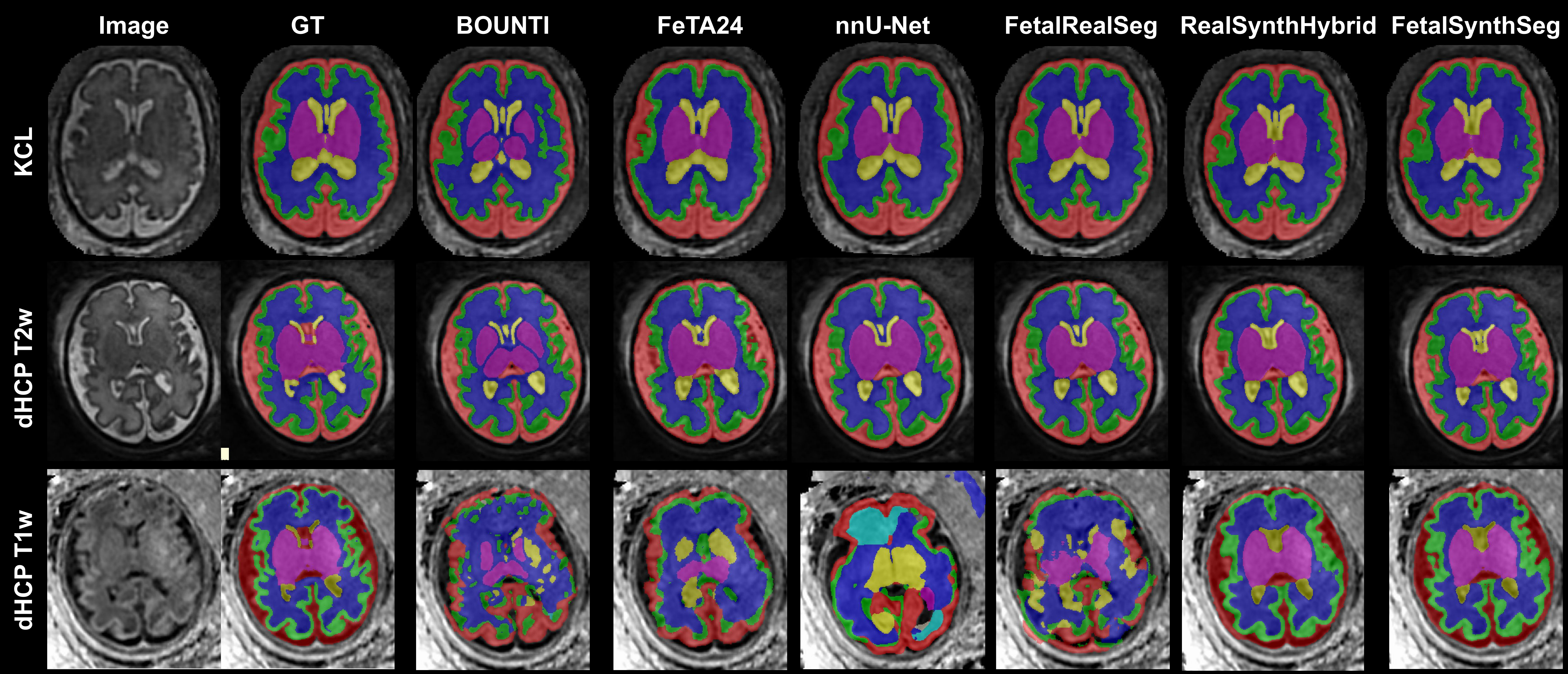}
    \caption{Qualitative comparison of the proposed models and \texttt{FeTA24} on a test case from the dHCP dataset, showing T2w images (top two rows) and T1w images (bottom two rows) from the same subject. White arrows highlight discrepancies between the dHCP and FeTA annotation schemes—particularly in the definitions of SGM and LV —as all models were trained using FeTA annotations, while the ground truth reflects dHCP labels converted to FeTA format. Yellow arrows indicate prominent segmentation errors made by the model fine-tuned on real T2w images when applied to the T1w contrast.}
    \label{fig:dhcp_ext}
\end{figure*}

\textcolor{black}{On dHCP~T2w, BOUNTI attains the highest performance: this could be explained by a closer alignment between the dHCP and BOUNTI labels (the latter were refined based on the dHCP annotations) as well as by a data leakage problem, as BOUNTI's training data included some of the dHCP~T2w data (though the exact data splits of dHCP used for BOUNTI training were not published). In contrast, \frs and \fetseg reach performance comparable to nnU-Net without such overlap, highlighting the strength of the proposed training schemes.}  Additional HD95 and per-class results are provided in Table \ref{tab:t2t1_side_by_side} in the Supplementary materials, which show that all methods exhibit similar relative performance patterns across labels, despite differing in their absolute metric values.

\begin{figure}[t]
\centering

\includegraphics[width=1\linewidth]{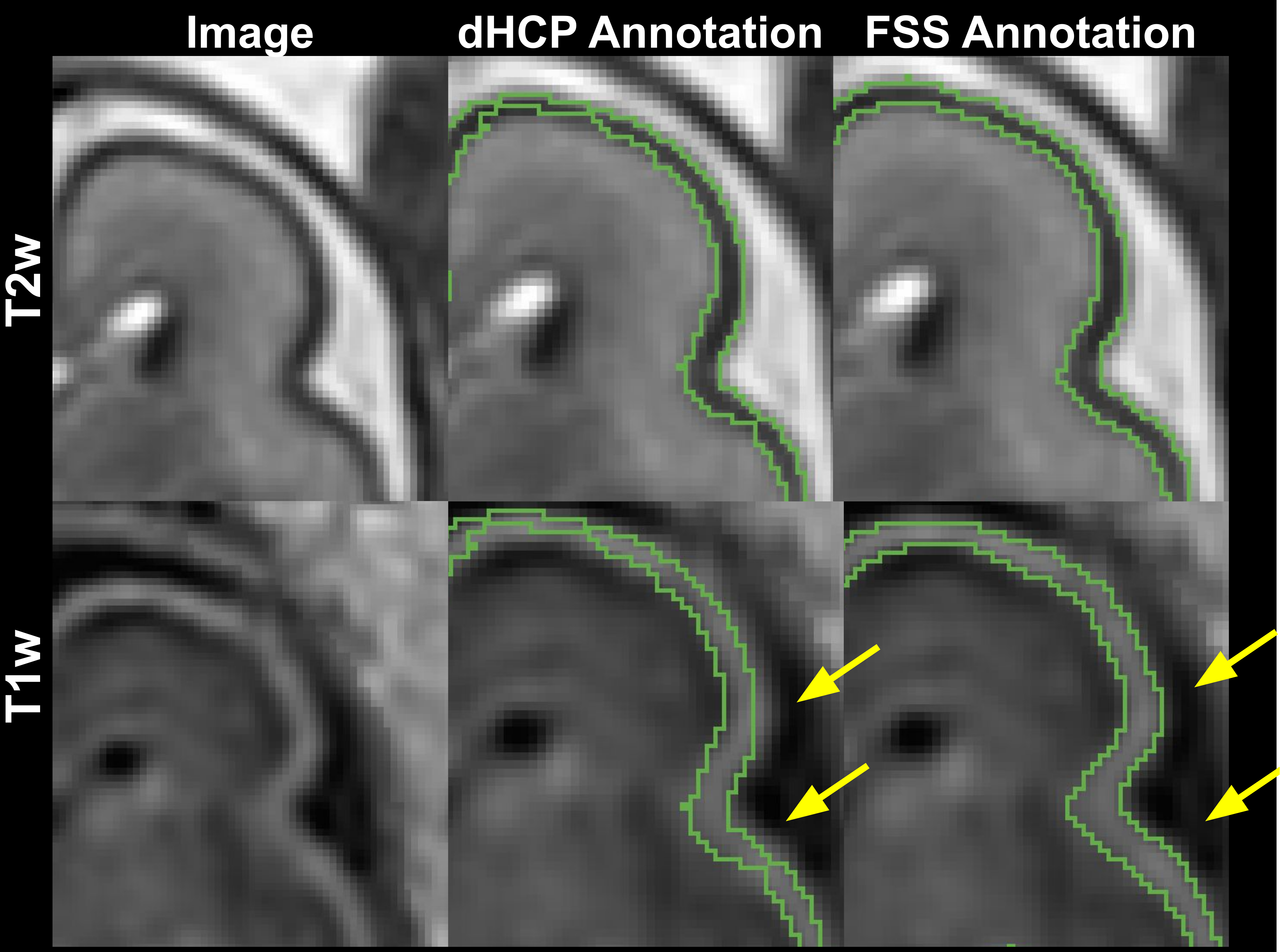}
\vspace{-.6cm}
\caption{Cortical GM segmentations on T1w and T2w dHCP images. dHCP Annotation: reference annotations obtained via registration of T2w-based labels to T1w. Bottom: \fetseg predictions obtained independently for each modality using the same model, showing improved alignment with modality-specific tissue boundaries.}
\label{fig:registrationerrordetection}
\end{figure}

The most striking effect appears in the OOD evaluation on dHCP~T1w. Here, \fetseg achieves a Dice of 80.1\%, while all other methods, including \frs, nnU-Net, BOUNTI, and FeTA24, drop below 25\%. \fetseg is thus the only evaluated model capable of reliably segmenting this challenging modality using a single, contrast-agnostic network trained exclusively on synthetic data. Figure~\ref{fig:dhcp_ext} illustrates these findings qualitatively: models trained on real T2w data struggle on T1w images, whereas \fetseg preserves anatomical plausibility and label consistency.
 \textcolor{black}{Besides a quantitative evaluation, visual inspection also shows that \fetseg produces highly consistent segmentations across T1w and T2w images. On T1w images, a visual assessment shows that these segmentations improve the quality of cortical gray matter segmentations compared to reference annotations obtained from propagating T2w labels on the T1w images, which are often imprecise, as illustrated on Figure~\ref{fig:registrationerrordetection}.}

\subsubsection*{Failure mode analysis}

\begin{figure}[t]
\centering
\includegraphics[width=1\linewidth]{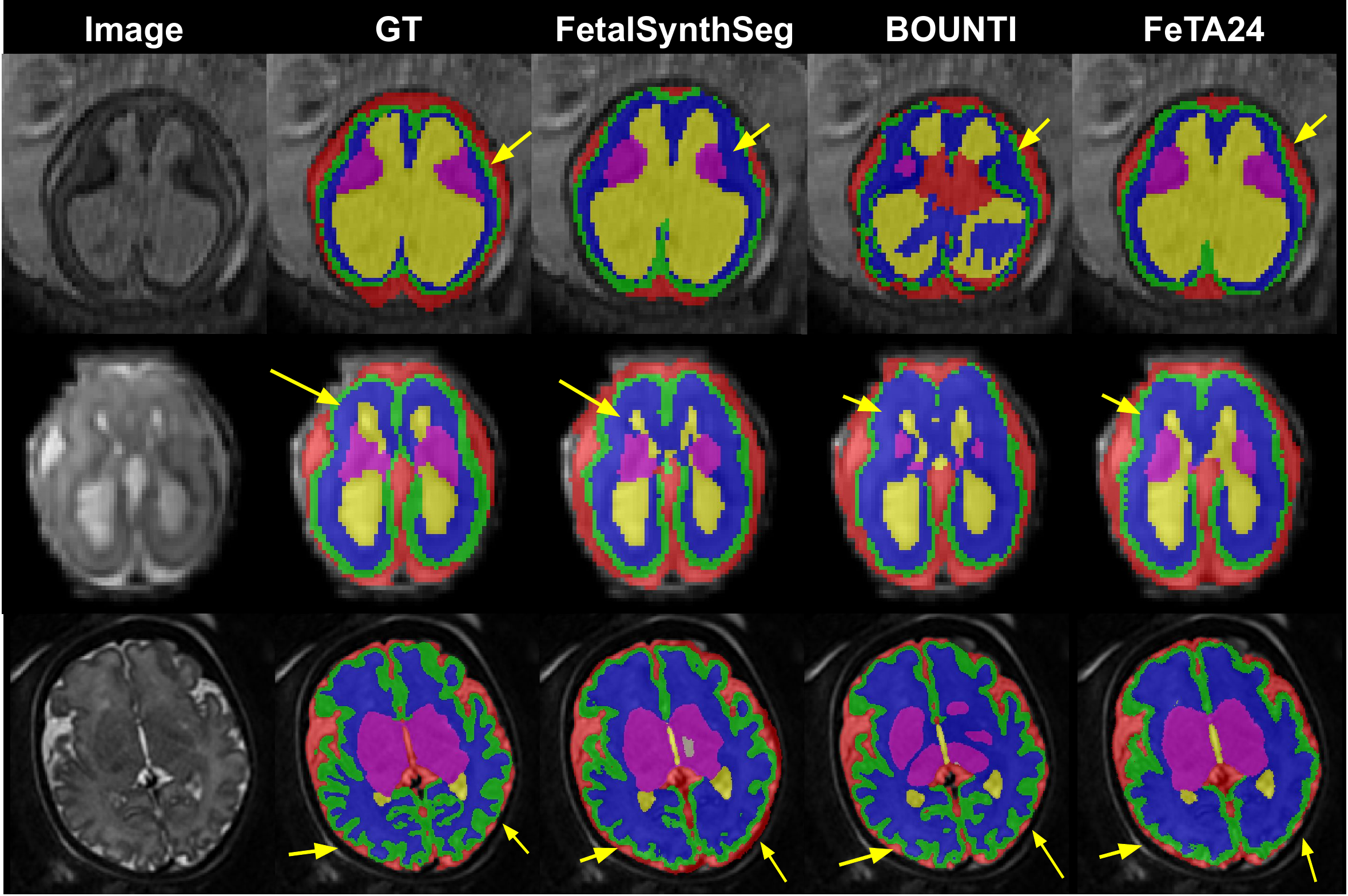}
\caption{\textcolor{black}{Representative failure cases of \fetseg compared to two state-of-the-art methods (BOUNTI and FeTA24). \textbf{Top row:} severely pathological case (ventriculomegaly), with errors highlighted in subcortical gray matter (yellow arrows). \textbf{Middle row:} very young subject with annotation errors due to challenging manual segmentation (yellow arrow indicates ventricle over-segmentation in the ground truth). \textbf{Bottom row:} older gestational age subject, where cortical gray matter segmentation is overly smooth (yellow arrows).}}
\label{fig:fss_failure}

\includegraphics[width=1\linewidth]{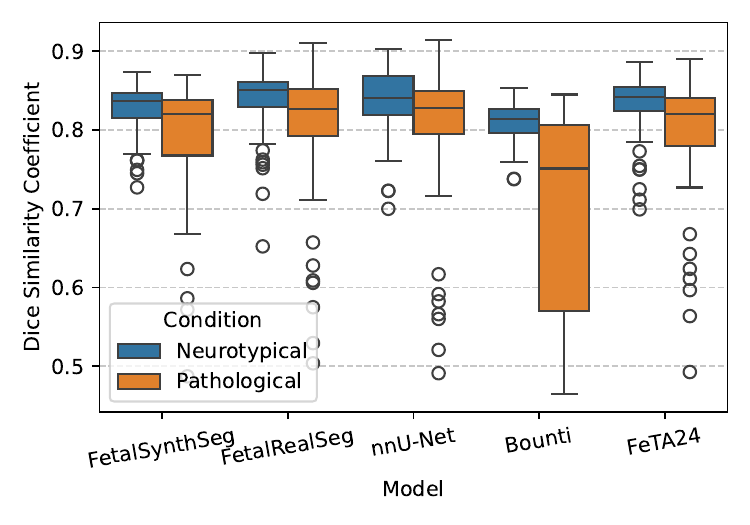}
\caption{\textcolor{black}{Dice score distributions of selected state-of-the-art methods on CHUV and KCL datasets, stratified by healthy and pathological subjects. Pathological cases show lower median performance and higher variability across all methods. All methods except for BOUNTI trained on the FeTA 2024 training data.}}
\label{fig:conditionperformance}
\end{figure}

\textcolor{black}{When analyzing the failure cases of \fetseg, three main categories emerge, as illustrated in Fig.~\ref{fig:fss_failure}.}

\textcolor{black}{First, \textbf{severely pathological cases} remain challenging for all methods. As shown in Fig.~\ref{fig:conditionperformance}, segmentation performance degrades and variance increases for pathological subjects, particularly in cases with strong anatomical alterations (e.g., ventriculomegaly; Fig.~\ref{fig:fss_failure}, top row). Notably, both \fetseg and the FeTA 2024 model outperform BOUNTI in this regime. This can be attributed to the limited presence of extreme anatomical abnormalities in BOUNTI’s training data~\citep{Uus2023.04.18.537347_Bounti}, in contrast to the FeTA dataset, which contains a substantial proportion of pathological cases, including severe abnormalities such as spina bifida and ventriculomegaly.}

\textcolor{black}{Second, reduced performance can often be attributed to \textbf{imperfect ground-truth annotations}. Strong partial volume effect and small structure sizes make manual annotation particularly challenging, and automated methods may produce anatomically more consistent segmentations than the provided ground truth. This is reflected in our results approaching or even exceeding the inter-rater variability reported for the FeTA dataset~\citep{zalevskyi2025advancesautomatedfetalbrain}.}

\textcolor{black}{Finally, \fetseg shows reduced performance in \textbf{older fetuses}, particularly in structures undergoing significant changes during late gestation, such as cortical gray matter. In these cases, segmentations tend to be overly smooth and deviate from the ground truth (Fig.~\ref{fig:fss_failure}, bottom row). This effect is especially pronounced in the dHCP dataset, which contains a higher proportion of older subjects that lie outside the gestational age range represented in the FeTA training data. Retraining the model with a larger proportion of older subjects should however address this issue.}

\subsection{Prospective validation.}\label{exp:4} 
\textcolor{black}{Finally, we  carried out a prospective validation on multi-TE data acquired at 0.55~T at KCL, and reconstructed using the method of \citet{bulut2025physics}. The main challenge in segmenting these data lies in the contrast change across echo times, and algorithms often fail at higher echo times, where the tissue across contrast is low.}

\begin{figure}[t]
\centering
\includegraphics[width=1.\linewidth]{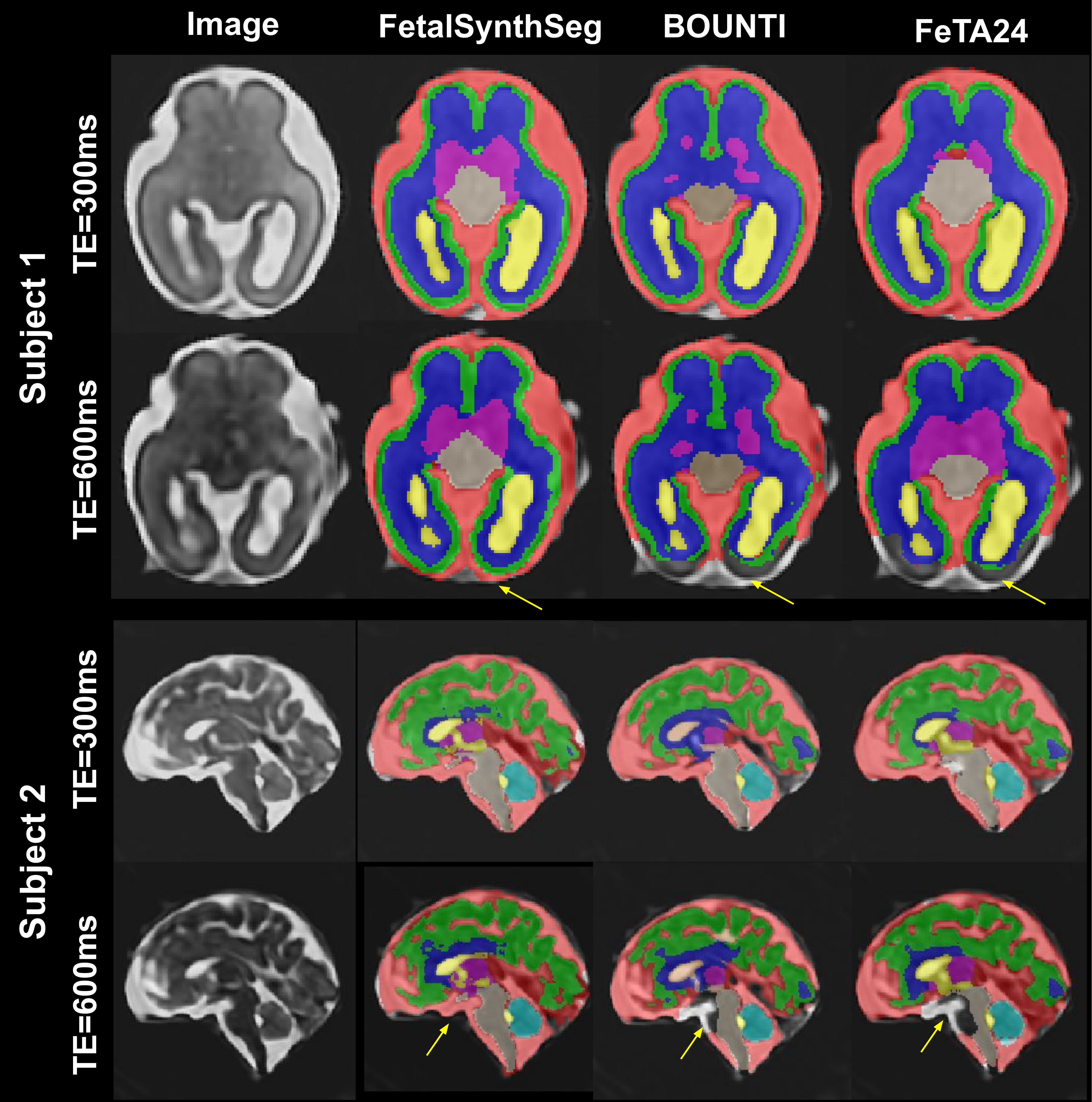}
\caption{\textbf{Example of TE-invariant segmentation for two subjects.} Two rows show images of the same subject acquired at different TEs. Columns show segmentations from different methods. \fetseg provides more consistent segmentations across TEs, supporting reliable T2 mapping.}
\label{fig:teconsistency}
\end{figure}

\textcolor{black}{As shown in Figure~\ref{fig:teconsistency}, \fetseg delivers more consistent segmentations across echo times, especially at higher TEs, where BOUNTI and FeTA24 fail to segment the occipital lobe (subject 1), as well as the brainstem (especially the pons) and CSF around it (subject 2). This is confirmed by quantitative results: since no ground-truth annotations are available for these low-field acquisitions, we compute cross-TE metrics. As the Shapiro-Wilk test for normality does not reject the null hypothesis, we compare the results using a paired t-test. The results show that \fetseg achieves a DSC of 0.89±0.02 compared to 0.84±0.04 (BOUNTI; p=0.018) and 0.82±0.02 (FeTA24; p=0.038), while also obtaining a lower HD95 (9.18±5.20 vs. 12.5±0.30 and 13.20±2.57, respectively), although in this case the results are not statistically significant.}


\section{Discussion}
In this work, we systematically analyzed single-source domain generalization for fetal brain tissue segmentation, building on \synthseg-style domain randomization~\citep{billot2021synthseg,billot2023synthseg}. Using more than 300 subjects across multiple sites, field strengths, reconstruction methods, and modalities, we disentangled the roles of intensity simulation, clustering, meta-classes, and real-image training. \textcolor{black}{We also confirmed both of our hypotheses: domain randomization is indeed more efficient that physics-based simulation for SSDG, and it enables strong OOD generalization while retaining a competitive ID performance.} Below, we summarize the main findings.

\noindent\textbf{(i) Intensity clustering is essential for strong domain randomization.}
Introducing intensity-based clustering within labels yielded an average improvement of $\sim$5 Dice points and was critical for enabling \synthseg-based models to outperform, on average, models trained solely on real data. Clustering increases intra-class variability and better captures subtle fetal tissue heterogeneity, particularly when only few anatomical labels are available. Consistent with observations in cardiac MRI~\citep{billot2023robust} and recent extensions~\citep{valabregue2024comprehensive}, our results confirm that structured intensity diversification is a key ingredient for robust SSDG.

\noindent\textbf{(ii) Meta-classes improve the trade-off between realism and randomization.}
Aggregating labels into meta-classes (WM, GM, CSF, non-brain) and then sub-clustering them, as done in \fetseg, consistently provided a small but statistically significant improvement over \synthseg ($\sim$2 Dice points). This design balances flexibility and prior knowledge: tissues expected to share similar intensities (e.g., CSF in ventricles and extra-axial spaces) are modeled jointly rather than forced to diverge, avoiding unnecessary use of capacity and yielding more coherent yet still highly varied synthetic contrasts.

\noindent\textbf{(iii) Physics-based simulators underperform randomized GMM-based generation for SSDG.}
Using \fabian~\citep{Lajous2022} with realistic T1/T2 values did not translate into strong cross-domain performance. Randomizing its parameters (\randfabian) improved results (to 68.4 Dice on average), but remained below \synthseg (72.2 Dice) and \fetseg (74.9 Dice). Direct GMM-based sampling more easily spans the large diversity of real-world contrasts than perturbations of physically grounded parameters alone. These findings suggest that, for SSDG under strong domain shifts, broad and sometimes non-physical intensity randomization is more beneficial than faithful but constrained physical simulation.

\noindent\textcolor{black}{\textbf{(iv) Training with real data is sufficient for some applications.}
While \fetseg's performance remains strong in-domain, its generalization ability comes at a small price: its performance in-domain is often slightly below models trained using purely real data, as was noticed in \citet{billot2023synthseg}. When evaluated on T2w datasets, \fetseg's DSC and HD95 is often slightly below \frs, \hrs and \texttt{nnU-Net} (the latter using an ensemble of models). The hybrid training of \hrs provides another layer of trade-offs: it is weaker in-domain compared to \frs, worse than \fetseg out-of-domain, but much more robust than \frs out-of-domain. Nevertheless, for all cases, an essential component lies in the data augmentation strategies used to train \fetseg and \frs (\ref{sec:aug}): they allow a stronger generalization than MONAI-based augmentations in every setting, which warrants further works to study these augmentations strategies more rigorously.}

\noindent\textbf{(v) \fetseg enables new applications in fetal brain MRI.}
\fetseg shows robust generalization across sites, SR methods, field strengths, and, importantly, contrasts. Trained exclusively on synthetic data, it:
\begin{enumerate}[leftmargin=.8cm,parsep=-2pt,itemsep=1pt]
    \item Matches or closely approaches strong real-data baselines on in-domain T2w datasets;
    \item Is, to our knowledge, the only evaluated model providing reliable segmentation on fetal dHCP T1w dataset (80\% Dice, versus $<25\%$ for all other methods);
    \item Maintains consistency across varying echo times (TEs) in T2-mapping acquisitions.
These properties allow \fetseg to act as a contrast-invariant backbone for: inter-modality registration support and QC, detection of label misregistrations (Fig.~\ref{fig:registrationerrordetection}),
and stable tissue delineation in quantitative mapping (Fig.~\ref{fig:teconsistency}).
\end{enumerate}

\textcolor{black}{We hypothesize that the performance gains achieved by \fetseg primarily arise from the proposed data generation and augmentation strategies, as the network architecture, loss function, and training procedure were kept unchanged throughout our experiments. In principle, these improvements should therefore be largely model-agnostic and transferable to other segmentation frameworks. Consequently, \fetseg data generation strategy could potentially be combined with alternative architectures, including more recent transformer-based models, ensembles, or different optimization and training strategies, which may further improve performance. However, validating this generalizability across architectures and training paradigms warrants further investigation.}

\paragraph{Limitations and future work.} Our experiments focus on fetal brain MRI only, a domain with pronounced shifts and limited labels. While this makes it an informative stress-test for SSDG, extrapolation to other anatomies or modalities should be done cautiously. Evaluations on additional public datasets and non-fetal applications are a natural next step.

Our work deliberately fixed a simple 3D U-Net backbone to isolate the effects of data generation. Architectural choices, ensembling, or topology-aware losses could interact with synthetic data strategies in non-trivial ways, and may further boost performance.

Another limitation relies on hand-crafted generators, GMM-based intensity sampling with user-defined ranges. More expressive generative models (e.g., conditional or diffusion-based approaches) could in principle learn richer priors over anatomy, artifacts, and contrasts. However, their computational cost and complexity currently hinder large-scale, on-the-fly generation~\citep{wang2023patch}, which is central to our efficiency-focused SSDG setting. Exploring hybrid schemes that combine learned generative models with lightweight randomization is an important direction for future work.

\textcolor{black}{Although our study includes a diverse set of datasets, most are available only after site-specific SRR, due to privacy constraints and the lack of access to raw low-resolution data. As a result, each dataset is typically associated with a single SRR pipeline, introducing a potential confounding factor. This makes it difficult to disentangle the respective contributions of acquisition domain and reconstruction method to the observed segmentation performance. A natural direction for future work is to systematically reconstruct datasets using multiple SRR methods and explicitly evaluate their impact on segmentation methods.}

\section{Conclusions}
We investigated how to best leverage synthetic data for fetal brain tissue segmentation under single-source domain generalization. Comparing physics-based, purely randomized, and hybrid generation strategies, we found that:
\begin{itemize}[leftmargin=.8cm,parsep=-2pt,itemsep=1pt]
    \item Structured intensity clustering and meta-class design are crucial for effective domain randomization;
    \item Broad GMM-based randomization outperforms physically constrained simulations for handling strong domain shifts;
\end{itemize}

Guided by these insights, we introduced \fetseg, a \synthseg-based framework tailored to fetal MRI. \fetseg attains competitive performance on in-domain T2w data and, notably, strong out-of-domain generalization across sites, reconstruction methods, and contrasts—including accurate segmentation of T1w fetal brain MRI using only synthetic training data.

Overall, our results demonstrate that carefully designed domain randomization, grounded in intensity clustering and meta-class modeling, provides a practical and efficient path toward contrast- and protocol-robust fetal brain segmentation in data-scarce settings.


\acks{This research was funded by the Swiss National Science Foundation (182602 and 215641),
ERA-NET Neuron MULTI-FACT project (SNSF 31NE30 203977), UKRI FLF\newline
(MR/T018119/1) and DFG Heisenberg funding (502024488); we acknowledge the
Leenaards and Jeantet Foundations as well as CIBM Center for Biomedical Imaging, a Swiss research center of excellence founded and supported by CHUV, UNIL, EPFL, UNIGE and HUG. This research was also supported by grants from NVIDIA and utilized NVIDIA RTX6000 ADA GPUs.

The Developing Human Connectome Project (dHCP) was funded by the European Research Council (ERC) under the European Union’s Seventh Framework Programme (FP7/2007–2013), Grant Agreement No. 319456. The Wellcome Centre for Integrative Neuroimaging is supported by core funding from the Wellcome Trust  [Grant No.\newline 203139/Z/16/Z].
The views expressed in this work are those of the authors and do not necessarily reflect those of the NHS, the NIHR, King’s College London, or the Department of Health and Social Care.
The authors gratefully acknowledge the families and participants involved in the dHCP, as well as the neonatal staff at the Evelina Newborn Imaging Centre, St Thomas’ Hospital, Guy’s \& St Thomas’ NHS Foundation Trust, London, UK, for their work in acquiring and processing the data.}

%
\ethics{The work follows appropriate ethical standards in conducting research and writing the manuscript, following all applicable laws and regulations regarding treatment of animals or human subjects.}

\coi{The author discloses receiving an NVIDIA Academic Hardware Grant, through which NVIDIA supplied GPU hardware that supported the experiments reported in this manuscript.}

\data{The KISPI and dHCP  are publicly available dataset \citep{FeTA_Synapse, Edwards2022}. KCL and CHUV datasets used for evaluation in this study are part of the private testing set from the FeTA challenge and are not publicly accessible. VIEN data is not publicly available.}

\bibliography{sample}


\clearpage
\onecolumn

\appendix

\makeatletter
\setcounter{table}{0}
\setcounter{figure}{0}
\renewcommand 
\thesection{S\@arabic\c@section}
\renewcommand\thetable{S\@arabic\c@table}
\renewcommand \thefigure{S\@arabic\c@figure}
\makeatother

\noindent\textbf{\Large Appendix}\\[2mm] 
\textbf{Table of contents:}
\begin{table}[h]
    \centering
    \begin{tabular}{lp{12cm}}
     \ref{sec:aug} (p. \pageref{sec:aug}) & Ablation on data augmentations \\
     \ref{sec:fabian} (p. \pageref{sec:fabian})& Comparing \fabian vs \fetseg on the same number of images\\
     \ref{sec:gen_param} (p. \pageref{sec:gen_param})& Generation parameters\\
     \ref{sec:quant} (p. \pageref{sec:quant}) & {Additional quantitative results}\\
     \ref{sec:oodres} (p. \pageref{sec:oodres}) & Out-of-domain evaluation and SoTA comparison\\
\end{tabular}
\end{table}

\section{Ablation on data augmentations}\label{sec:aug}
In this additional experiment, we compared the \synthseg-based augmentations to a common set of augmentations from MONAI (referred to as \textit{simple aug.}). We re-trained the models using the following augmentations, each applied with a random probability of 0.5: random affine deformation (rotation=0.2, scale=0.1, translate=30, sheer=0.1), random contrast change (Gamma transformation) ($\gamma\in [0.5, 1.5]$), random Gaussian noise ($\mu=0, \sigma=0.1$), random blurring ($\sigma\in[0.5, 1.5]$), and scaling to the 0-1 range. Compared to these \textit{simple aug.}, \synthseg-based augmentation has three additional components: a random non-linear deformation, random bias field, and a random resampling simulating an acquisition at a different resolution.

The results in Table~\ref{tab:2_results_fssVSALL_supp} show a different picture than the results in the main paper. When comparing models trained using simple augmentations, the model trained using real data outperforms \fetseg (69.9 vs 66.9 Dice). Moving from simple augmentation increases the performance of all methods considered: \fabian (+2.9), \randfabian (+3.9), \fetseg (+8.0), \texttt{Real Data} (+1.9). This illustrates clearly how these data augmentation strategies are instrumental to the performance of \fetseg. They are also strong contenders for a model trained on real data and could be used in the future. A detailed ablation study of augmentations would be an interesting future step.

\begin{table*}[h]

\centering
\caption{Mean Dice for models trained on different sources of data. The variance is computed over all testing subjects within a split. The best performing method is shown in \textbf{boldface} in each column, and the second best is \underline{underlined}. An asterisk ($^*$) indicates that the best method is statistically significantly better than the second best (Wilcoxon test with correction).}
\label{tab:2_results_fssVSALL_supp}
\resizebox{\linewidth}{!}{%
\begin{tblr}{
  column{3} = {c},
  column{4} = {c},
  column{5} = {c},
  column{6} = {c},
  column{7} = {c},
  column{8} = {c},
  column{9} = {c},
  cell{1}{1} = {r=2}{},
  cell{1}{3} = {c=2}{},
  cell{1}{5} = {c=2}{},
  cell{1}{7} = {c=2}{},
  cell{3}{1} = {r=3}{},
  cell{6}{1} = {r=4}{},
  hline{1,12} = {-}{0.08em},
  hline{2} = {2}{},
  hline{2} = {3-8}{l},
  hline{3} = {2-9}{0.03em},
  hline{6,10} = {-}{0.05em},
}
{\textbf{Augmentation}\\\textbf{type}} & \textbf{Testing split}  & \textbf{CHUV-MIAL}    &                       & \textbf{KISPI-IRTK}   &                       & \textbf{KISPI-MIAL}    &                        & \textbf{Global}        \\
                                       & \textbf{Training split} & KISPI-IRTK            & KISPI-MIAL            & CHUV-MIAL             & KISPI-MIAL            & CHUV-MIAL              & KISPI-IRTK             &                        \\
Simple                                 & FaBiAN                  & 66.9\,$\pm$\,5.1      & 70.3\,$\pm$\,6.9      & 51.6\,$\pm$\,14.1     & 56.7\,$\pm$\,14.2     & 59.8\,$\pm$\,15.2      & 57.3\,$\pm$\,17.9      & 60.4\,$\pm$\,14.3      \\
                                       & randFaBiAN              & 74.1\,$\pm$\,3.1      & 69.0\,$\pm$\,6.4      & 63.8\,$\pm$\,10.3     & 54.0\,$\pm$\,9.3      & 62.0\,$\pm$\,13.8      & 62.8\,$\pm$\,13.0      & 64.3\,$\pm$\,12.1      \\
                                       & FetalSynthSeg           & 72.9\,$\pm$\,4.0      & 70.3\,$\pm$\,6.2      & 71.4\,$\pm$\,9.3      & 68.3\,$\pm$\,8.2      & 58.7\,$\pm$\,15.5      & 60.0\,$\pm$\,16.1      & 66.9\,$\pm$\,11.8      \\
SynthSeg                               & FaBiAN                  & 74.2\,$\pm$\,4.2      & 73.1\,$\pm$\,5.7      & 53.6\,$\pm$\,13.3     & 56.6\,$\pm$\,12.9     & 60.6\,$\pm$\,17.1      & 61.5\,$\pm$\,20.1      & 63.3\,$\pm$\,15.5      \\
                                       & randFaBiAN              & \uline{79.1\,$\pm$\,2.3}  & \uline{78.2\,$\pm$\,2.9}  & 55.1\,$\pm$\,12.7     & 68.9\,$\pm$\,7.8      & 60.7\,$\pm$\,7.3       & \uline{68.1\,$\pm$\,14.7}  & 68.3\,$\pm$\,13.8      \\
                                       & SynthSeg                & 75.9\,$\pm$\,3.9      & 73.7\,$\pm$\,3.9      & \uline{70.9\,$\pm$\,9.2}  & \uline{74.8\,$\pm$\,7.8}  & 60.5\,$\pm$\,15.7      & 63.4\,$\pm$\,16.8      & 72.2\,$\pm$\,13.0      \\
                                       & FetalSynthSeg           & $\textbf{80.7\,$\pm$\,2.0}^*$ & 76.9\,$\pm$\,3.3      & $\textbf{79.2\,$\pm$\,9.0}^*$ & $\textbf{76.8\,$\pm$\,6.9}^*$ & \uline{67.5\,$\pm$\,16.0}  & $\textbf{68.5\,$\pm$\,15.6}$ & $\textbf{74.9\,$\pm$\,11.5}^*$ \\
Simple                                 & Real Data               & 76.5\,$\pm$\,3.2      & 75.2\,$\pm$\,3.5      & 69.6\,$\pm$\,13.6     & 67.7\,$\pm$\,12.9     & 67.2\,$\pm$\,17.6      & 63.4\,$\pm$\,17.0      & 69.9\,$\pm$\,12.6      \\
SynthSeg                               & Real Data               & 77.2\,$\pm$\,4.0      & $\textbf{78.5\,$\pm$\,3.3}^*$ & 70.6\,$\pm$\,13.9     & 71.9\,$\pm$\,11.5     & $\textbf{68.0\,$\pm$\,19.3}$ & 64.3\,$\pm$\,19.1      & \uline{71.8\,$\pm$\,13.9}  
\end{tblr}
}
\end{table*}

\newpage
\section{Comparing \fabian vs \fetseg on the same number of images}\label{sec:fabian}
To ensure a fair comparison between \fabian and \fetseg we equalize the computational budget used to generate images for these two approaches. We generate offline 6000 synthetic images per split, as in our \fabian experiments, and use them to train a \texttt{FetalSynthSeg-6k} model. Table \ref{tab:fss6k} shows that although this model achieves slightly lower results than \fetseg, it follows similar trends in performance across splits and still outperforms even the \randfabian approach overall (average Dice of \texttt{FetalSynthSeg-6k} 72.2±8.8 is vs average Dice of \randfabian of 68.3±8.5 across all splits).

\begin{table*}[htb]
\centering
\caption{Dice scores comparing \fetseg model trained on the same amount of synthetic images as \fabian models.}
\label{tab:fss6k}
\resizebox{.9\linewidth}{!}{%
\begin{tblr}{
  row{1} = {c},
  row{2} = {c},
  cell{1}{2} = {c=2}{},
  cell{1}{4} = {c=2}{},
  cell{1}{6} = {c=2}{},
  cell{1}{8} = {r=2}{},
  hline{1,7} = {-}{0.08em},
  hline{2} = {1-7}{},
  hline{3} = {-}{},
}
\textbf{Testing Split} & \textbf{CHUV-MIAL} & & \textbf{KISPI-IRTK} & & \textbf{KISPI-MIAL} & & \textbf{Global} \\
\textbf{Training Split} & KISPI-IRTK & KISPI-MIAL & CHUV-MIAL & KISPI-MIAL & CHUV-MIAL & KISPI-IRTK & \\
\fabian & 74.2±4.2 & 73.1±5.7 & 53.6±13.3 & 56.6±12.9 & 60.6±17.1 & 61.5±20.1 & 63.3±15.5 \\
\randfabian & \underline{79.1±2.3} & \textbf{78.2±2.9*} & 55.1±12.7 & 68.9±7.8 & 60.7±7.3 & \underline{68.1±14.7} & 68.3±13.8 \\
\texttt{FetalSynthSeg-6k} & 79.0±2.6 & 76.0±3.6 & \underline{78.3±9.2} & \underline{75.1±6.6} & \underline{66.5±15.9} & 67.7±15.4 & \underline{72.2±13.0} \\
\fetseg & \textbf{80.7±2.0$^{*}$} & \underline{76.9±3.3} & \textbf{79.2±9.0$^{*}$} & \textbf{76.8±6.9$^{*}$} & \textbf{67.5±16.0$^{*}$} & \textbf{68.5±15.6$^{*}$} & \textbf{74.9±11.5*} \\
\end{tblr}
}
\end{table*}
\section{Generation parameters}\label{sec:gen_param}
In the Table \ref{table:haste_params} we report parameters used to create synthetic images with the FaBiAN generator. 
\begin{table}[h!]
\centering
\small
\begin{tabular}{ll}
\toprule
\textbf{Contrast}               &                                 \\ \midrule
Effective echo time (ms)        & {[}90,300{]}                    \\ 
Echo spacing (ms)               & 6.12                            \\ 
Echo train length               & 150                             \\ 
Excitation flip angle (°)       & 90                              \\ 
Refocusing pulse flip angle (°) & {[}150,180{]}                   \\ \midrule
\textbf{Geometry}               &                                 \\ \midrule 
Slice thickness (mm)            & 0.8                             \\ 
Slice gap (mm)                  & 0                               \\ 
Number of slices                & 112                             \\ 
Phase oversampling (\%)         & 80                              \\ 
Shift of field-of-view (mm)     & 0                               \\ \midrule
\textbf{Resolution}             &                                 \\ \midrule
Field-of-view (mm\textsuperscript{2}) & 120x120                    \\ 
Base resolution (voxels)        & 150                             \\ 
Phase resolution (\%)           & 70                              \\ 
Reconstruction matrix           & 150                             \\ 
Zero-interpolation filling      & 1                               \\ \midrule
\textbf{Acceleration technique} &                                 \\ \midrule
Reference lines                 & 42                              \\ 
Acceleration factor             & 2                               \\ \midrule
\textbf{Noise}                  &                                 \\ \midrule
Mean                            & 0                               \\ 
Standard deviation              & \textless{}0.01                 \\ \bottomrule
\end{tabular}
\caption{\fabian Simulation Parameters for HASTE Imaging}
\label{table:haste_params}
\end{table}
\newpage
\section{Additional quantitative results}\label{sec:quant}
We present additional quantitative results related to the conducted experiments, reporting the Dice scores per tissue for all of the tested models in Tables \ref{tab:2fetalsynthseg_comparative_full} as well as the average 95th-percentile Hausdorff distance score in Table \ref{tab:hdtable}. 

\begin{table*}[h!]
\caption{Mean and standard deviation of Dice score of the explored models for different testing and training splits. LV - lateral ventricles, CBM - cerebellum, SGM - sub-cortical gray matter, BS - brainstem, Mean DSC - value averaged across all segmentation labels.}
\label{tab:2fetalsynthseg_comparative_full}
\resizebox{\linewidth}{!}{
\begin{tblr}{
  row{1} = {c},
  cell{2}{1} = {r=10}{c},
  cell{2}{2} = {r=5}{c},
  cell{7}{2} = {r=5}{c},
  cell{12}{1} = {r=10}{c},
  cell{12}{2} = {r=5}{c},
  cell{17}{2} = {r=5}{c},
  cell{22}{1} = {r=10}{c},
  cell{22}{2} = {r=5}{c},
  cell{27}{2} = {r=5}{c},
  hline{1,32} = {-}{0.08em},
  hline{2,12,22} = {-}{},
  hline{7,17,27} = {2-11}{},
}
{\textbf{ Testing}\\\textbf{~Splits}} & {\textbf{ Training}\\\textbf{~Split}} & \textbf{ Experiment} & \textbf{Mean DSC } & \textbf{CSF } & \textbf{GM } & \textbf{WM } & \textbf{LV } & \textbf{CBM } & \textbf{SGM } & \textbf{BS }\\
\begin{sideways}CHUV-MIAL\end{sideways} & \begin{sideways}KISPI-IRTK\end{sideways} & FetalSynthSeg & \textbf{80.7$\pm$2.0} & \textbf{79.3$\pm$2.7} & \textbf{70.7$\pm$3.6} & \textbf{86.4$\pm$2.3} & \textbf{$78.7\pm 5.4$} & \textbf{$88.5\pm 2.3$} & \textbf{79.6$\pm$5.4} & \textbf{81.9$\pm$2.7}\\
 &  & Real data & 77.2$\pm$4.0 & 72.3$\pm$7.1 & 68.4$\pm$4.7 & 86.0$\pm$2.8 & 75.8$\pm$7.9 & 84.3$\pm$9.2 & 76.4$\pm$5.9 & 77.5$\pm$4.7\\
 &  & FaBiAN & 74.2$\pm$4.2 & 75.3$\pm$4.2 & 64.2$\pm$4.8 & 81.3$\pm$4.4 & 71.9$\pm$9.9 & 85.3$\pm$3.6 & 62.8$\pm$14.2 & 78.7$\pm$3.3\\
 &  & randFaBiAN & 79.1$\pm$2.3 & 77.0$\pm$3.3 & 69.7$\pm$3.6 & 85.9$\pm$2.3 & 75.7$\pm$5.3 & 87.4$\pm$3.1 & 78.5$\pm$8.4 & 79.5$\pm$3.9\\
 &  & SynthSeg & 75.9$\pm$3.9 & 73.7$\pm$4.5 & 64.1$\pm$6.3 & 83.1$\pm$4.4 & 71.9$\pm$8.5 & 85.8$\pm$2.9 & 74.0$\pm$10.6 & 78.8$\pm$3.9\\
 & \begin{sideways}KISPI-MIAL\end{sideways} & FetalSynthSeg & 76.9$\pm$3.3 & 73.1$\pm$8.1 & 60.5$\pm$4.8 & 84.9$\pm$3.0 & \textbf{77.3$\pm$6.3} & 86.2$\pm$4.3 & 83.7$\pm$4.6 & 72.8$\pm$3.9\\
 &  & Real data & \textbf{78.5$\pm$3.3} & \textbf{79.0$\pm$5.5} & \textbf{65.6$\pm$4.4} & \textbf{85.8$\pm$3.2} & 77.1$\pm$7.3 & 86.5$\pm$6.1 & \textbf{83.9$\pm$3.8} & 71.4$\pm$5.1\\
 &  & FaBiAN & 73.1$\pm$5.7 & 73.3$\pm$4.9 & 61.9$\pm$5.9 & 81.1$\pm$4.8 & 66.3$\pm$10.6 & 81.0$\pm$14.9 & 76.3$\pm$7.3 & 71.5$\pm$7.1\\
 &  & randFaBiAN & 78.2$\pm$2.9 & 77.8$\pm$3.9 & 64.1$\pm$5.2 & 84.4$\pm$3.2 & 75.5$\pm$5.4 & \textbf{87.5$\pm$4.6} & 83.0$\pm$5.0 & \textbf{75.3$\pm$5.0}\\
 &  & SynthSeg & 73.7$\pm$4.4 & 67.9$\pm$12.0 & 55.1$\pm$8.7 & 81.3$\pm$4.2 & 72.1$\pm$7.2 & 84.0$\pm$3.0 & 80.9$\pm$5.2 & 74.6$\pm$3.9\\
\begin{sideways}KISPI-IRTK\end{sideways} & \begin{sideways}CHUV-MIAL\end{sideways} & FetalSynthSeg & \textbf{79.2$\pm$9.0} & \textbf{77.6$\pm$10.9} & \textbf{69.4$\pm$12.6} & \textbf{85.0$\pm$12.9} & \textbf{79.7$\pm$5.7} & \textbf{87.5$\pm$9.1} & \textbf{74.8$\pm$14.4} & \textbf{80.3$\pm$6.6}\\
 &  & Real data & 70.6$\pm$13.9 & 58.6$\pm$18.4 & 60.9$\pm$15.9 & 84.8$\pm$9.8 & 72.8$\pm$11.7 & 74.3$\pm$26.3 & 74.5$\pm$11.0 & 68.1$\pm$19.9\\
 &  & FaBiAN & 53.6$\pm$13.3 & 51.4$\pm$18.0 & 40.9$\pm$11.9 & 74.9$\pm$10.7 & 36.0$\pm$13.8 & 45.9$\pm$27.0 & 66.9$\pm$12.1 & 59.0$\pm$19.3\\
 &  & randFaBiAN & 55.1$\pm$12.7 & 54.2$\pm$17.2 & 41.5$\pm$11.5 & 76.9$\pm$10.5 & 37.2$\pm$13.7 & 51.8$\pm$28.0 & 66.5$\pm$11.7 & 57.8$\pm$18.5\\
 &  & SynthSeg & 70.9$\pm$9.2 & 69.9$\pm$11.6 & 58.6$\pm$9.0 & 79.2$\pm$11.8 & 66.4$\pm$7.0 & 80.2$\pm$11.3 & 68.8$\pm$13.7 & 73.2$\pm$9.2\\
 & \begin{sideways}KISPI-MIAL\end{sideways} & FetalSynthSeg & \textbf{76.8$\pm$6.9} & \textbf{78.7$\pm$9.2} & \textbf{68.5$\pm$9.1} & \textbf{86.5$\pm$10.9} & \textbf{80.8$\pm$4.7} & 85.1$\pm$9.1 & 68.2$\pm$12.3 & 69.9$\pm$6.8\\
 &  & Real data & 71.9$\pm$11.5 & 72.3$\pm$19.5 & 65.3$\pm$13.4 & 85.5$\pm$7.6 & 71.6$\pm$9.0 & 80.3$\pm$20.2 & 64.8$\pm$13.3 & 63.2$\pm$14.5\\
 &  & FaBiAN & 56.6$\pm$12.9 & 58.0$\pm$17.6 & 39.3$\pm$13.0 & 74.3$\pm$9.5 & 50.6$\pm$15.0 & 64.7$\pm$27.7 & 58.7$\pm$13.9 & 50.6$\pm$18.7\\
 &  & randFaBiAN & 68.9$\pm$7.8 & 60.8$\pm$16.0 & 52.8$\pm$11.1 & 82.2$\pm$6.3 & 68.1$\pm$9.7 & 80.1$\pm$10.9 & 68.0$\pm$12.3 & 70.2$\pm$7.9\\
 &  & SynthSeg & 74.8$\pm$7.8 & 74.4$\pm$12.7 & 65.7$\pm$10.8 & 85.4$\pm$11.0 & 71.6$\pm$7.3 & \textbf{86.3$\pm$7.5} & \textbf{68.5$\pm$13.1} & \textbf{72.0$\pm$8.0}\\
\begin{sideways}KISPI-MIAL\end{sideways} & \begin{sideways}CHUV-MIAL\end{sideways} & FetalSynthSeg & 67.5$\pm$16.0 & 59.5$\pm$28.3 & 48.7$\pm$19.4 & 74.3$\pm$17.3 & \textbf{78.6$\pm$12.3} & \textbf{67.2$\pm$29.7} & \textbf{79.3$\pm$9.8} & \textbf{64.9$\pm$16.0}\\
 &  & Real data & \textbf{68.0$\pm$19.3} & \textbf{62.1$\pm$32.4} & \textbf{58.3$\pm$15.7} & \textbf{81.8$\pm$12.3} & 77.0$\pm$14.5 & 64.6$\pm$37.3 & 78.8$\pm$9.9 & 53.4$\pm$29.8\\
 &  & FaBiAN & 60.6$\pm$17.1 & 52.9$\pm$29.1 & 48.7$\pm$15.2 & 77.6$\pm$12.3 & 63.2$\pm$17.3 & 53.7$\pm$33.1 & 73.4$\pm$14.5 & 54.8$\pm$25.8\\
 &  & randFaBiAN & 60.7$\pm$16.3 & 52.3$\pm$28.6 & 49.6$\pm$14.5 & 78.2$\pm$12.4 & 63.7$\pm$18.0 & 52.8$\pm$32.5 & 73.8$\pm$14.3 & 54.6$\pm$22.4\\
 &  & SynthSeg & 60.5$\pm$15.7 & 53.7$\pm$26.0 & 43.8$\pm$14.1 & 70.0$\pm$15.4 & 65.9$\pm$12.1 & 59.9$\pm$28.5 & 71.1$\pm$15.5 & 59.0$\pm$17.0\\
 & \begin{sideways}KISPI-IRTK\end{sideways} & FetalSynthSeg & \textbf{68.5$\pm$15.6} & \textbf{63.2$\pm$25.9} & \textbf{57.2$\pm$16.8} & 82.1$\pm$13.5 & 79.4$\pm$12.2 & \textbf{73.2$\pm$24.9} & \textbf{63.6$\pm$18.4} & \textbf{60.8$\pm$17.0}\\
 &  & Real data & 64.3$\pm$19.1 & 59.8$\pm$29.0 & 55.0$\pm$17.9 & 79.1$\pm$15.5 & 72.6$\pm$12.9 & 64.4$\pm$33.8 & 62.5$\pm$16.3 & 57.1$\pm$21.6\\
 &  & FaBiAN & 61.5$\pm$20.1 & 57.5$\pm$31.5 & 53.1$\pm$19.0 & 80.6$\pm$11.8 & 72.3$\pm$16.1 & 59.1$\pm$35.1 & 52.8$\pm$20.9 & 54.9$\pm$23.4\\
 &  & randFaBiAN & 68.1$\pm$14.7 & 62.5$\pm$27.3 & 59.4$\pm$12.4 & \textbf{83.8$\pm$12.2} & \textbf{80.4$\pm$11.0} & 70.6$\pm$25.0 & 62.3$\pm$21.2 & 57.4$\pm$15.1\\
 &  & SynthSeg & 63.4$\pm$16.8 & 57.3$\pm$26.8 & 50.2$\pm$17.6 & 77.4$\pm$16.0 & 77.7$\pm$13.4 & 70.4$\pm$22.0 & 54.7$\pm$23.9 & 56.2$\pm$18.2
\end{tblr}
}
\end{table*}

\begin{table*}[h!]
\caption{Mean and standard deviation of 95-th percentile Hasudorff distance for the explored models on different testing and training splits. LV - lateral ventricles, CBM - cerebellum, SGM - sub-cortical gray matter, BS - brainstem, mHD95 - average value across all labels.}
\label{tab:hdtable}
\resizebox{\linewidth}{!}{

\begin{tblr}{
  row{1} = {c},
  cell{2}{1} = {r=10}{c},
  cell{2}{2} = {r=5}{c},
  cell{7}{2} = {r=5}{c},
  cell{12}{1} = {r=10}{c},
  cell{12}{2} = {r=5}{c},
  cell{17}{2} = {r=5}{c},
  cell{22}{1} = {r=10}{c},
  cell{22}{2} = {r=5}{c},
  cell{27}{2} = {r=5}{c},
  hline{1,32} = {-}{0.08em},
  hline{2,12,22} = {-}{},
  hline{7,17,27} = {2-11}{},
}
{\textbf{Testing}\\\textbf{Splits}} & {\textbf{ Training}\\\textbf{Split}} & \textbf{ Experiment} & \textbf{mHD95 } & \textbf{CSF } & \textbf{GM } & \textbf{WM } & \textbf{Ventricles } & \textbf{Cerebellum } & \textbf{Deep\_GM } & \textbf{Brainstem }\\
\begin{sideways}CHUV - MIAL\end{sideways} & \begin{sideways}KISPI-IRTK\end{sideways} & FetalSynthSeg & \textbf{2.0±0.5} & \textbf{1.7±0.3} & \textbf{1.6±0.3} & 2.5±3.3 & \textbf{2.0±0.9} & \textbf{1.5±0.2} & \textbf{2.8±0.6} & \textbf{2.1±0.6}\\
 &  & Real data & 2.9±1.2 & 2.6±0.8 & 1.9±0.5 & \textbf{2.0±0.3} & 4.2±4.4 & 2.9±4.7 & 4.0±2.8 & 2.8±1.1\\
 &  & FaBiAN & 3.6±1.3 & 2.6±0.9 & 2.2±0.4 & 3.6±0.8 & 5.3±4.1 & 3.5±6.0 & 5.0±2.3 & 2.9±1.0\\
 &  & randFaBiAN & 2.1±0.3 & 1.8±0.3 & \textbf{1.6±0.3} & 2.1±0.3 & 2.1±0.7 & 1.6±0.4 & 3.0±0.8 & 2.6±1.0\\
 &  & SynthSeg & 2.7±0.6 & 2.0±0.4 & 2.1±1.9 & 2.5±0.3 & 3.7±1.8 & 2.3±0.8 & 3.3±0.8 & 2.9±1.4\\
 & \begin{sideways}KISPI-MIAL\end{sideways} & FetalSynthSeg & 3.1±0.7 & 2.7±1.7 & 2.5±0.5 & \textbf{2.2±0.4} & 2.6±1.7 & 2.9±2.9 & \textbf{2.6±0.5} & 6.3±0.9\\
 &  & Real data & 3.0±0.7 & 2.2±1.6 & 2.3±0.4 & \textbf{2.2±0.5} & 3.1±3.0 & \textbf{1.9±0.8} & 2.9±0.6 & 6.6±1.1\\
 &  & FaBiAN & 4.1±1.4 & 3.1±1.9 & 2.8±0.6 & 4.0±1.2 & 4.6±3.4 & 2.8±1.7 & 5.7±5.2 & 5.4±1.3\\
 &  & randFaBiAN & \textbf{2.7±0.6} & \textbf{1.9±0.6} & \textbf{2.2±0.5} & 2.4±0.5 & \textbf{2.1±0.6} & 2.5±2.2 & 2.8±0.7 & 5.0±1.2\\
 &  & SynthSeg & 3.4±0.6 & 3.5±2.9 & 2.4±0.4 & 2.9±0.3 & 4.6±1.9 & 3.1±1.2 & 3.2±0.6 & \textbf{3.9±0.9}\\
\begin{sideways}KISPI - IRTK\end{sideways} & \begin{sideways}CHUV-MIAL\end{sideways} & FetalSynthSeg & 2.5±1.6 & 2.4±2.2 & 1.6±0.8 & 2.2±1.4 & 1.9±1.0 & 1.5±0.9 & 4.3±4.0 & 3.4±3.6\\
 &  & Real data & 9.8±5.1 & 9.0±2.3 & 7.6±4.3 & 5.7±4.2 & 6.3±6.0 & 16.1±16.1 & 9.5±9.6 & 14.1±9.7\\
 &  & FaBiAN & 14.8±5.3 & 9.2±2.1 & 10.3±2.2 & 9.6±2.7 & 14.2±4.3 & 28.7±16.6 & 15.1±8.7 & 16.3±11.1\\
 &  & randFaBiAN & 14.8±5.1 & 9.0±2.0 & 10.7±2.3 & 10.0±3.2 & 13.2±5.5 & 24.4±16.5 & 17.1±10.1 & 18.9±11.3\\
 &  & SynthSeg & \textbf{1.5±0.5} & \textbf{1.3±1.0} & \textbf{1.2±0.3} & \textbf{1.4±0.4} & \textbf{1.4±0.6} & \textbf{1.3±0.8} & \textbf{2.0±0.8} & \textbf{1.7±0.7}\\
 & \begin{sideways}KISPI-MIAL\end{sideways} & FetalSynthSeg & \textbf{3.4±1.7} & \textbf{2.4±1.5} & \textbf{1.7±0.8} & \textbf{2.1±0.7} & \textbf{3.1±3.4} & \textbf{3.4±6.5} & 5.5±2.9 & 5.9±1.4\\
 &  & Real data & 5.9±4.0 & 6.3±4.0 & 4.1±4.8 & 5.4±6.8 & 4.3±4.8 & 6.3±10.8 & 6.8±4.2 & 7.8±4.7\\
 &  & FaBiAN & 14.0±5.0 & 9.0±2.1 & 10.1±2.7 & 11.1±3.4 & 8.9±6.3 & 24.9±18.4 & 18.6±11.0 & 15.4±9.1\\
 &  & randFaBiAN & 5.5±2.6 & 6.8±2.6 & 3.9±1.9 & 3.8±2.3 & 5.3±5.6 & 6.7±10.0 & 6.3±5.6 & 5.8±3.2\\
 &  & SynthSeg & 3.5±1.4 & 2.9±2.0 & \textbf{1.7±0.8} & 2.3±0.7 & 3.8±3.2 & 3.7±6.3 & \textbf{5.3±2.0} & \textbf{4.4±1.3}\\
\begin{sideways}KISPI - MIAL\end{sideways} & \begin{sideways}CHUV-MIAL\end{sideways} & FetalSynthSeg & 4.1±2.4 & 4.8±4.5 & 3.2±1.9 & 3.4±1.8 & 2.4±2.3 & 4.5±3.9 & 3.8±3.3 & 6.4±3.3\\
 &  & Real data & 5.1±3.4 & 5.0±4.9 & 3.1±1.7 & 2.8±1.3 & 4.1±4.1 & 7.7±9.8 & 4.1±3.0 & 8.6±6.3\\
 &  & FaBiAN & 8.2±4.3 & 7.3±4.9 & 6.5±2.4 & 4.1±1.4 & 8.9±4.4 & 11.2±12.4 & 7.3±7.7 & 12.3±8.7\\
 &  & randFaBiAN & 7.6±4.2 & 7.0±4.8 & 6.0±1.8 & 3.9±1.4 & 8.0±4.4 & 9.9±10.6 & 7.8±8.1 & 10.5±8.1\\
 &  & SynthSeg & \textbf{2.6±1.6} & \textbf{3.1±3.2} & \textbf{2.2±2.2} & \textbf{2.6±3.0} & \textbf{1.7±1.0} & \textbf{2.5±1.7} & \textbf{2.7±3.4} & \textbf{3.6±4.3}\\
 & \begin{sideways}KISPI-IRTK\end{sideways} & FetalSynthSeg & \textbf{3.9±2.0} & \textbf{3.9±3.6} & \textbf{2.4±1.5} & \textbf{2.6±1.3} & 2.3±2.0 & 3.6±3.1 & \textbf{5.6±2.6} & 6.7±2.7\\
 &  & Real data & 4.7±2.7 & 5.2±4.6 & 2.8±1.7 & 2.7±1.4 & 4.7±3.9 & 5.9±6.5 & 5.8±2.6 & \textbf{5.9±1.8}\\
 &  & FaBiAN & 6.5±3.9 & 6.5±4.9 & 4.7±3.3 & 4.4±2.4 & 6.4±5.4 & 8.3±7.9 & 7.0±5.5 & 8.4±6.0\\
 &  & randFaBiAN & 4.1±2.0 & 4.4±4.3 & 2.8±1.3 & 3.1±2.8 & \textbf{2.2±1.7} & \textbf{3.5±2.3} & 5.7±2.8 & 7.0±2.6\\
 &  & SynthSeg & 5.2±2.8 & 4.9±4.2 & 3.6±2.0 & 3.7±3.0 & 4.4±6.5 & 5.1±8.1 & 6.7±3.9 & 7.9±2.9
\end{tblr}
}
\end{table*}

\newpage~\newpage

\newpage
\section{Out-of-domain evaluation and comparison to state of the art}
\label{sec:oodres}

Table~\ref{tab:t2t1_side_by_side} reports the label-wise DSC and HD95 metrics for all evaluated methods, separated by modality (T1w and T2w). The results are averaged across the corresponding out-of-domain datasets to provide a robust comparison against state-of-the-art (SoTA) fetal brain MRI segmentation models.

On the \textbf{T1w datasets}, we observe a substantial performance gap between \fetseg and all competing methods. \fetseg is the only model that consistently performs well across \emph{all} labels, whereas other SoTA approaches show highly variable performance and generally fail to generalize to T1-weighted data. Notably, WM is the only label where some methods achieve moderate performance (with DSC values around 60\%), yet this remains far below the 89\% achieved by \fetseg. All other labels exhibit even larger drops. These results highlight that domain shift affects different anatomical structures unequally, with some tissues (e.g., WM) being significantly more robust to cross-domain variability than others.

On the \textbf{T2w datasets}, the performance across models is more homogeneous. For each label, the differences between SoTA methods typically remain within 1--3\% DSC, indicating that all methods generalize reasonably well to unseen T2-weighted data. Nevertheless, consistent label-specific trends are still present: WM is generally the easiest structure to segment, whereas VM and GM tend to be more challenging across all models. This reinforces that label-dependent difficulty persists even in settings with smaller domain gaps.

\begin{table*}[h]
\centering

\begin{minipage}[t]{0.48\textwidth}
\centering
\scriptsize
\begin{tabular}{l l c c}
\toprule
\multicolumn{4}{c}{\textbf{T2w datasets}} \\
\midrule
\textbf{Model} & \textbf{Label} & \textbf{DSC [\%]} & \textbf{HD95} \\
\midrule

\multirow{7}{*}{Bounti}
  & BS  & 88.0{\tiny±6.0}  & 2.7{\tiny±2.6} \\
  & CBM & 89.0{\tiny±2.0}  & 2.3{\tiny±2.3} \\
  & CSF & 88.0{\tiny±8.0}  & 1.6{\tiny±1.4} \\
  & GM  & 81.0{\tiny±9.0}  & 1.2{\tiny±0.8} \\
  & SGM & 69.0{\tiny±8.0}  & 5.9{\tiny±1.6} \\
  & VM  & 77.0{\tiny±9.0}  & 7.9{\tiny±3.5} \\
  & WM  & 88.0{\tiny±10.0} & 3.4{\tiny±0.8} \\
\midrule
 
\multirow{7}{*}{FRS}
  & BS  & 83.0{\tiny±5.0}  & 4.2{\tiny±1.4} \\
  & CBM & 89.0{\tiny±5.0}  & 2.6{\tiny±1.0} \\
  & CSF & 85.0{\tiny±6.0}  & 1.9{\tiny±2.1} \\
  & GM  & 79.0{\tiny±6.0}  & 1.6{\tiny±1.8} \\
  & SGM & 81.0{\tiny±6.0}  & 4.8{\tiny±1.5} \\
  & VM  & 78.0{\tiny±7.0}  & 8.6{\tiny±4.9} \\
  & WM  & 91.0{\tiny±3.0}  & 2.0{\tiny±2.1} \\
\midrule
 
\multirow{7}{*}{FSS}
  & BS  & 81.0{\tiny±5.0}  & 4.2{\tiny±1.6} \\
  & CBM & 90.0{\tiny±6.0}  & 2.2{\tiny±1.0} \\
  & CSF & 84.0{\tiny±9.0}  & 1.8{\tiny±1.1} \\
  & GM  & 78.0{\tiny±9.0}  & 1.4{\tiny±0.5} \\
  & SGM & 80.0{\tiny±7.0}  & 4.8{\tiny±1.7} \\
  & VM  & 77.0{\tiny±9.0}  & 6.9{\tiny±4.5} \\
  & WM  & 90.0{\tiny±4.0}  & 1.8{\tiny±0.4} \\
\midrule
 
\multirow{7}{*}{FeTA24}
  & BS  & 81.0{\tiny±6.0}  & 4.0{\tiny±1.5} \\
  & CBM & 87.0{\tiny±6.0}  & 2.5{\tiny±1.0} \\
  & CSF & 83.0{\tiny±8.0}  & 2.3{\tiny±1.9} \\
  & GM  & 77.0{\tiny±7.0}  & 1.6{\tiny±0.8} \\
  & SGM & 80.0{\tiny±8.0}  & 4.5{\tiny±1.7} \\
  & VM  & 77.0{\tiny±9.0}  & 6.6{\tiny±4.8} \\
  & WM  & 90.0{\tiny±4.0}  & 2.0{\tiny±0.7} \\
\midrule
 
\multirow{7}{*}{nnU-Net}
  & BS  & 81.0{\tiny±8.0}  & 4.3{\tiny±1.6} \\
  & CBM & 89.0{\tiny±8.0}  & 2.5{\tiny±1.1} \\
  & CSF & 84.0{\tiny±9.0}  & 2.3{\tiny±3.9} \\
  & GM  & 80.0{\tiny±7.0}  & 1.4{\tiny±0.6} \\
  & SGM & 81.0{\tiny±8.0}  & 4.7{\tiny±1.7} \\
  & VM  & 80.0{\tiny±8.0}  & 7.0{\tiny±4.9} \\
  & WM  & 91.0{\tiny±3.0}  & 1.8{\tiny±1.9} \\
\bottomrule
\end{tabular}
\end{minipage}
\hfill
\begin{minipage}[t]{0.48\textwidth}
\centering
\scriptsize
\begin{tabular}{l l c c}
\toprule
\multicolumn{4}{c}{\textbf{T1w datasets}} \\
\midrule
\textbf{Model} & \textbf{Label} & \textbf{DSC [\%]} & \textbf{HD95} \\
\midrule
 
\multirow{7}{*}{Bounti}
  & BS  & 0.0{\tiny±1.0}   & 32.3{\tiny±9.5} \\
  & CBM & 2.0{\tiny±4.0}   & 36.3{\tiny±11.2} \\
  & CSF & 21.0{\tiny±3.0}  & 5.3{\tiny±1.1} \\
  & GM  & 8.0{\tiny±3.0}   & 6.5{\tiny±1.4} \\
  & SGM & 19.0{\tiny±11.0} & 13.4{\tiny±3.1} \\
  & VM  & 5.0{\tiny±3.0}   & 10.8{\tiny±3.2} \\
  & WM  & 60.0{\tiny±6.0}  & 7.3{\tiny±1.3} \\
\midrule
 
\multirow{7}{*}{FRS}
  & BS  & 51.0{\tiny±12.0} & 11.1{\tiny±6.8} \\
  & CBM & 63.0{\tiny±24.0} & 7.5{\tiny±7.9} \\
  & CSF & 31.0{\tiny±6.0}  & 6.3{\tiny±4.1} \\
  & GM  & 22.0{\tiny±5.0}  & 5.4{\tiny±4.2} \\
  & SGM & 58.0{\tiny±10.0} & 12.2{\tiny±7.3} \\
  & VM  & 15.0{\tiny±8.0}  & 13.4{\tiny±7.8} \\
  & WM  & 67.0{\tiny±4.0}  & 7.8{\tiny±6.9} \\
\midrule
 
\multirow{7}{*}{FSS}
  & BS  & 78.0{\tiny±3.0}  & 5.5{\tiny±0.6} \\
  & CBM & 89.0{\tiny±2.0}  & 2.7{\tiny±0.8} \\
  & CSF & 79.0{\tiny±6.0}  & 2.0{\tiny±0.5} \\
  & GM  & 75.0{\tiny±5.0}  & 1.6{\tiny±0.6} \\
  & SGM & 80.0{\tiny±4.0}  & 5.8{\tiny±0.5} \\
  & VM  & 69.0{\tiny±7.0}  & 10.3{\tiny±2.0} \\
  & WM  & 89.0{\tiny±3.0}  & 1.9{\tiny±0.5} \\
\midrule
 
\multirow{7}{*}{FeTA24}
  & BS  & 5.0{\tiny±8.0}   & 16.5{\tiny±5.8} \\
  & CBM & 33.0{\tiny±18.0} & 10.7{\tiny±6.0} \\
  & CSF & 28.0{\tiny±4.0}  & 5.4{\tiny±1.7} \\
  & GM  & 15.0{\tiny±4.0}  & 5.2{\tiny±1.6} \\
  & SGM & 11.0{\tiny±11.0} & 17.9{\tiny±5.6} \\
  & VM  & 7.0{\tiny±3.0}   & 12.8{\tiny±3.4} \\
  & WM  & 65.0{\tiny±6.0}  & 5.7{\tiny±2.1} \\
\midrule
 
\multirow{7}{*}{nnU-Net}
  & BS  & 0.0{\tiny±0.0}   & 41.3{\tiny±8.2} \\
  & CBM & 2.0{\tiny±7.0}   & 41.1{\tiny±10.5} \\
  & CSF & 29.0{\tiny±4.0}  & 10.8{\tiny±8.7} \\
  & GM  & 9.0{\tiny±4.0}   & 14.2{\tiny±9.1} \\
  & SGM & 1.0{\tiny±2.0}   & 25.4{\tiny±10.4} \\
  & VM  & 3.0{\tiny±2.0}   & 27.4{\tiny±16.7} \\
  & WM  & 51.0{\tiny±10.0} & 21.0{\tiny±12.2} \\
\bottomrule
\end{tabular}
\end{minipage}
 
\caption{Segmentation performance on T2w (left) and T1w (right) datasets. DSC in percent, HD95 in original units (mean$\pm$std).}
\label{tab:t2t1_side_by_side}
\end{table*}

\end{document}